\documentclass[12pt,usenames,dvipsnames]{article}

\usepackage{latexsym}
\usepackage{amssymb,amsfonts,amsmath}
\usepackage{graphicx} 
\usepackage{indentfirst}
\usepackage{bbm}
\usepackage{amssymb}
\usepackage{verbatim}
\usepackage{amsmath, amsthm,amssymb}
\usepackage{mathrsfs}
\usepackage{hyperref}
\usepackage{amsfonts}
\usepackage{dsfont}
\usepackage{cite}
\usepackage{xcolor}
\usepackage{enumerate}
\usepackage{cleveref}

\parskip=\medskipamount

\newcommand {\cD}{{\cal D}}
\newcommand {\cE}{{\cal E}}

\newcommand {\cJ}{{\cal J}}
\newcommand {\cK}{{\cal K}}
\newcommand {\cL}{{\cal L}}
\newcommand {\cM}{{\cal M}}
\newcommand {\cN}{{\cal N}}
\newcommand {\cO}{{\cal O}}

\newcommand {\cQ}{{\cal Q}}

\def\a{\alpha}
\def \bi{\bibitem}

\def\b{\beta}
\def\c{\chi}
\def\d{\delta}
\def\e{\epsilon}
\def\f{\phi}
\def\g{\gamma}

\def\l{\lambda}
\def\m{\mu}
\def\n{\nu}
\def\o{\omega}
\def\p{\pi}
\def\q{\theta}
\def\r{\rho}
\def\s{\sigma}
\def\t{\tau}

\def\x{\xi}
\def\z{\zeta}
\def\D{\Delta}
\def\F{\Phi}
\def\J{\Psi}
\def\L{\Lambda}
\def\O{\Omega}
\def\P{\Pi}

\def\S{\Sigma}
\def\U{\Upsilon}
\def\X{\Xi}
\def\tr{{\rm tr}}
\def\rd{{\rm d}}
\def\ri{{\rm i}}
\def\re{{\rm e}}

\def\N{{\cal N}}

\newcommand{\ad}{{\dot{\alpha}}}                           
\newcommand{\bd}{{\dot{\beta}}}                            
\newcommand{\ve}{\varepsilon}                            
\newcommand{\cDB}{{\bar\cD}}                            

\newcommand{\ab}{{\a\b}}

\renewcommand{\aa}{{\a\ad}}
\newcommand{\bb}{{\b\bd}}
\newcommand{\pa}{\partial}                           
\newcommand{\hf}{\frac12}

\newcommand{\vf}{\varphi}

\newcommand{\be}{\begin{equation}}
\newcommand{\ee}{\end{equation}}
\newcommand{\bea}{\begin{eqnarray}}
\newcommand{\eea}{\end{eqnarray}}
\newcommand{\non}{\nonumber}
\newcommand{\bm}[1]{\mbox{\boldmath$#1$}}

\def\double #1{#1{\hbox{\kern-2pt $#1$}}}

\newcommand{\hal}{{\hat{\a}}}

\newcommand{\gd}{{\dot\g}}

\newif\ifdtup

\newcommand{\bsubeq}{\begin{subequations}}
\newcommand{\esubeq}{\end{subequations}}

\numberwithin{equation}{section}

\newcommand{\sSU}{\mathsf{SU}}
\newcommand{\sSL}{\mathsf{SL}}
\newcommand{\sGL}{\mathsf{GL}}
\newcommand{\sSO}{\mathsf{SO}}

\newcommand{\sU}{\mathsf{U}}

\newcommand{\sOSp}{\mathsf{OSp}}

\newcommand{\id}{\mathds{1}}

\newcommand{\dmu}{{\dot{\mu}}}

\newcommand{\ah}{{\hat{\a}}}

\newcommand{\T}{\text{T}}
\newcommand{\sT}{\text{sT}}
\newcommand{\braket}[2]{\langle #1 | #2 \rangle}

\begin{document}

\begin{titlepage}
\begin{flushright}
December 2025
\end{flushright}
\vspace{5mm}

\begin{center}
{\Large \bf 
Anti-de Sitter flag superspace
}
\end{center}

\begin{center}

{\bf Nowar E. Koning and Sergei M. Kuzenko} \\
\vspace{5mm}

\footnotesize{ 
{\it Department of Physics M013, The University of Western Australia\\
35 Stirling Highway, Perth W.A. 6009, Australia}}  
~\\
\vspace{2mm}
~\\
Email: \texttt{nowar.koning@research.uwa.edu.au,
sergei.kuzenko@uwa.edu.au}\\
\vspace{2mm}

\end{center}

\begin{abstract}
\baselineskip=14pt
This work aims to develop a global formulation for ${\cal N}=2$ harmonic/projective anti-de Sitter (AdS) superspace $\text{AdS}^{4|8}\times S^2   \simeq \text{AdS}^{4|8}\times {\mathbb C}P^1$ that allows for a simple action of superconformal (and hence AdS isometry) transformations. First of all, we provide an alternative supertwistor description of the ${\cal N}$-extended AdS superspace in four dimensions, AdS$^{4|4\cal N}$, which corresponds to a realisation of the connected component $\mathsf{OSp}_0({\cal N}|4; {\mathbb R})$ of the AdS isometry supergroup as $\mathsf{SU}(2,2 |{\cal N}) \bigcap \mathsf{OSp} ({\cal N}| 4; {\mathbb C})$. 
The proposed realisation yields the following properties: (i) AdS$^{4|4\cal N}$ is an open domain  of the compactified ${\cal N}$-extended Minkowski superspace, $\overline{\mathbb M}^{4|4\cal N}$; (ii) the infinitesimal ${\cal N}$-extended superconformal transformations naturally act on AdS$^{4|4\cal N}$; and (iii) the isometry transformations 
of AdS$^{4|4\cal N}$ are described by those superconformal transformations which obey a certain constraint. 
 The obtained results for AdS$^{4|4\cal N}$  are then applied to develop a supertwistor  formulation for an AdS flag  superspace $ \text{AdS}^{4|8} \times {\mathbb F}_1(2)$ that we identify with the ${\cal N}=2$ harmonic/projective AdS superspace. This construction makes it possible to read off the superconformal and AdS isometry transformations acting on the analytic subspace of the harmonic superspace. 
\end{abstract}
\vspace{5mm}

\begin{flushright}
{\it Dedicated to Jim Gates on the occasion of his 75th birthday$\qquad{}$}
\end{flushright}
\vspace{5mm}

\vfill

\vfill
\end{titlepage}

\newpage
\renewcommand{\thefootnote}{\arabic{footnote}}
\setcounter{footnote}{0}

\tableofcontents{}
\vspace{1cm}
\bigskip\hrule

\allowdisplaybreaks

\section{Introduction}

As is well-known, there exist two fully-fledged superspace approaches to formulate off-shell $\cN=2$ rigid supersymmetric field theories in four dimensions: (i) harmonic superspace \cite{GIKOS,GIOS};  and (ii) projective superspace \cite{KLR,LR1,LR2}. They make use of the same superspace 
\bea
{\mathbb M}^{4|8}\times {\mathbb C}P^1  \simeq {\mathbb M}^{4|8}\times  \sSU(2)\big/\sU(1) 
\eea
which was introduced for the first time by Rosly \cite{Rosly}.
However, they differ in the following conceptual points: (i) the structure of off-shell   supermultiplets used; 
and (ii) the supersymmetric action principle chosen.\footnote{The relationship between the harmonic and projective superspace formulations is spelled out in \cite{K_double, Jain:2009aj, Kuzenko:2010bd, Butter:2012ta}.} 
In particular, they deal with different off-shell realisations for the so-called charged hypermultiplet:
(i) the $q$-hypermultiplet \cite{GIKOS} in harmonic superspace; and (ii) the polar hypermultiplet \cite{LR1} in projective superspace.\footnote{The terminology ``polar hypermultiplet'' was introduced in the influential paper \cite{G-RRWLvU}.} 

In 2007, both the harmonic and projective superspace approaches were extended to the case of $\cN=1$ supersymmetric theories in AdS$_5$  \cite{Kuzenko:2007aj, Kuzenko:2007vs}. The projective superspace construction of 
\cite{Kuzenko:2007aj, Kuzenko:2007vs}, in conjunction with the concept of superconformal projective multiplets
\cite{Kuzenko:2006mv, Kuzenko:2007qy}, has proved to be powerful for nontrivial generalisations. 
It has been used for developing off-shell formulations for general supergravity-matter systems, first in five dimensions \cite{Kuzenko:2007cj, Kuzenko:2007hu, KT-M5D08}, and soon after in four
\cite{KLRT-M_4D-1, Kuzenko:2008qz, KLRT-M_4D-2}, three \cite{KLT-M11} and six \cite{Linch:2012zh} dimensions.
In a locally supersymmetric framework, the superspaces AdS$^{4|8}$ and AdS$^{5|8}$ originate as maximally supersymmetric solutions in the 4D $\cN=2$ \cite{BK10,BK11} and 5D $\cN=1$ \cite{Kuzenko:2014eqa} AdS supergravity theories obtained by coupling the corresponding Weyl multiplet to two conformal 
compensators: (i) the vector multiplet; and (ii) the $\cO(2)$ multiplet.  

Extending the covariant harmonic-superspace approach developed for AdS$_5$ \cite{Kuzenko:2007aj, Kuzenko:2007vs}, or  its four-dimensional  analogue introduced recently in \cite{Gargett:2025xcg}, to local supersymmetry has turned out to be a nontrivial technical problem. 
To explain this issue, it suffices to restrict our attention to the 4D $\cN=2$ case and consider the $\sSU(2)$ superspace formulation for $\cN=2$ conformal supergravity \cite{KLRT-M_4D-1}. Let $\cD_A = (\cD_a , \cD_\a^i, \bar \cD^\ad_i)$ be 
the corresponding covariant derivatives for curved superspace  $\cM^{4|8}$, and let $v^+_i$ be the homogeneous coordinates for ${\mathbb C}P^1$. 
The algebra of supergravity covariant derivatives \cite{KLRT-M_4D-1} implies that the 
the spinor operators $\cD^+_{ \a}:=v^+_i\cD^i_{ \a} $ and   
${\bar \cD}^+_{\dot  \a}:=v^+_i{\bar \cD}^i_{\dot \a} $ satisfy the anti-commutation relations:
\bea
\{  \cD^+_{ \a} , \cD^+_{ \b} \}
=4\, Y_{\a \b}\,{\mathbb J}^{++}
+4 \, S^{++}M_{\a \b}~, \qquad 
\{\cD_\a^+,\cDB_\bd^+\}=
8 \,G_{\a \bd} \,{\mathbb J}^{++}~,
\label{analyt}
\eea
with
${\mathbb J}^{++}:=v^+_i v^+_j {\mathbb J}^{ij}$ and 
$S^{++}:=v^+_i v^+_j S^{ij}$. Here $M_{\a\b}$ and ${\mathbb J}^{ij}$ are the Lorentz and $\sSU(2)$ generators,
while $Y_{\a\b}$, $G_{\a\bd}$ and $S^{ij}$ are torsion tensors. 
With the notation $\cD^+_{\hat \a} = (\cD^+_{ \a} , \bar \cD^+_{ \ad} )$,
a Grassmann analytic superfield
$Q$ is a  scalar superfield on curved superspace $\cM^{4|8}$ which is  $v$-dependent
and obeys the covariant Grassmann analyticity constraints 
\bea
\cD^+_{\hat \a} Q
=0
~.
\label{ana}
\eea
These constraints have  the integrability conditions $\{  \cD^+_{\hat \a} , \cD^+_{\hat \b} \} Q =0$. These conditions are automatically satisfied for the {\it projective multiplets}, which are characterised by the property  \cite{KLRT-M_4D-1} 
\bea
{\mathbb J}^{++} Q=0~.
\label{J++}
\eea
However, the integrability conditions do not hold for general {\it harmonic multiplets}.\footnote{This problem does not occur in the case of AdS superspace AdS$^{4|8}$ where $Y_{\a\b} =0$ and $G_{\a\bd}=0$ 
\cite{KT-M08}. The conditions $Y_{\a\b} =0$ and $G_{\a\bd}=0$ also hold in on-shell AdS supergravity upon imposing an appropriate  super-Weyl gauge \cite{BK10,BK11}. The superspace geometry of on-shell supergravity is determined by the chiral super-Weyl tensor $W_{\a\b}$ and the real iso-triplet $S^{ij}$.}

A covariant harmonic-superspace formulation for general $\cN=2$ supergravity-matter systems was developed  ten years ago by Butter \cite{ButterHSS}, who also presented a plethora of nontrivial applications. 
In his approach, the conventional harmonic superspace $\cM^{4|8}  \times S^2$ is replaced with 
$\cM^{4|8} \times T{\mathbb C}P^1$, where the internal space is the tangent bundle of ${\mathbb C}P^1$.
In the present paper we will advocate for a different internal space, namely a flag manifold ${\mathbb F}_1(2) $, which is often denoted $F(1, {\mathbb C}^2)$.
Instead of considering a generic $\cN=2$ curved superspace $\cM^{4|8}$, our attention will be restricted to the AdS case.
Our analysis applies to
the following AdS superspaces with auxiliary dimensions:
\begin{itemize}
\item $\cN=2$ AdS${}_4$ flag superspace\footnote{The flag superspaces associated with (complexified) $\cN$-extended Minkowski superspace were considered in \cite{Rosly,Rosly2,RoslyS, Howe:1995md}. In the case of $\cN=3$ supersymmetry, the relevant flag manifold is $F (1,2, {\mathbb C}^3) $ \cite{Rosly}, with its points 
being all possible sequences $V_1 \subset V_2 \subset {\mathbb C}^3$, where $V_1$ and $V_2$ are one- and two-dimensional subspaces of ${\mathbb C}^3$. Several important multiplets in $\cN=3$ conformal supergravity, including the super Bach tensor, are naturally defined on the manifold $\cM^{4|12} \times F (1,2, {\mathbb C}^3) $ \cite{Kuzenko:2023qkg}.}
\begin{subequations}
\bea
  \text{AdS}^{4|8} \times {\mathbb F}_1(2) ~, \qquad {\mathbb F}_1(2) = \sGL(2,{\mathbb C} ) \big/\widetilde{\mathbb H}_1(2) ~,
 \label{4D-AdS-flag} 
 \eea
 \item $\cN=1$ AdS${}_5$  flag superspace 
 \bea
  \text{AdS}^{5|8} \times {\mathbb F}_1(2) ~, \qquad {\mathbb F}_1(2) = \sGL(2,{\mathbb C} ) \big/\widetilde{\mathbb H}_1(2) ~,
 \label{5D-AdS-flag} 
 \eea
 \end{subequations} 
 \end{itemize}
 where $\widetilde{\mathbb H}_1(2) $ is the group of nonsingular lower triangular matrices, 
\bea
\widetilde{\mathbb H}_1(2) := \left\{ \tilde{\bm r}= \left(
\begin{array}{cc}
 a  ~& 0\\
b ~&      c 
\end{array}
\right) \in \sGL(2,{\mathbb C}) \right\}~.
\eea
Here the flag manifold ${\mathbb F}_1(2) $ is the space of flags $V_1 \subset V_2 = {\mathbb C}^2$, 
with $V_1$ a one-dimensional subspace of ${\mathbb C}^2$. Of course,  ${\mathbb F}_1(2) $ can be viewed 
as ${\mathbb C}P^1$ or as $S^2 \simeq  \sSU(2)\big/\sU(1) $, which are precisely the realisations corresponding to the projective and harmonic superspaces, respectively. However, its description as
$\sGL(2,{\mathbb C} ) \big/\widetilde{\mathbb H}_1(2)$ 
is most useful when dealing with $\cN=2$ superconformal transformations in the analytic subspace of harmonic superspace \cite{GIOS-conf}. 
An important fact is that the three equivalent realisations ${\mathbb F}_1(2) $ are naturally associated with different functional types of (super)fields. This point will be elaborated upon in Sections \ref{section2} and \ref{section3}
 which are devoted to the discussion of $\cN=2$ supersymmetric field theories on Minkowski flag superspace $ {\mathbb M}^{4|8} \times {\mathbb F}_1(2) $.
In the remainder of this paper we will concentrate on a global description of the flag superspace
\eqref{4D-AdS-flag} and develop its supertwistor realisation.

The supertwistor realisations for the $\cN$-extended AdS superspaces $\text{AdS}^{4|4\cal N}$ and $\text{AdS}^{5|8\cal N}$ were developed in Refs. \cite{Kuzenko:2021vmh, Koning:2023ruq, Koning:2024iiq} and \cite{Koning:2024vdl}, respectively. In the present paper (Sections \ref{section4} and \ref{section5}) we provide an alternative supertwistor description of AdS$^{4|4\cN}$, which corresponds to a realisation of the connected component
$\mathsf{OSp}_0({\cN}|4; {\mathbb R})$ of the AdS isometry supergroup 
 as $\mathsf{SU}(2,2 |{\cal N}) \bigcap \mathsf{OSp} ({\cal N}| 4; {\mathbb C})$.\footnote{It is well-known that the $\cN$-extended AdS superalgebra in four dimensions, $\mathfrak{osp}(\cN|4;{\mathbb R})$, is a subalgebra of the $\cN$-extended superconformal algebra  $\mathfrak{su}(2,2|\cN)$, see \cite{GGRS,Bandos:2002nn} as well as \cite{Ivanov:2025jdp} for a recent discussion.} 
 The advantage of doing so is that it allows us to read off the superconformal and isometry transformation rules 
  for AdS$^{4|4\cN}$ from those known for the compactified Minkowski superspace.
Section \ref{section6} is devoted to deriving a  supertwistor realisation of $ \text{AdS}^{4|8} \times {\mathbb F}_1(2) $.

The main body of the paper is accompanied by three technical appendices. Appendix \ref{appendixA} contains 
a brief review of $\cN=2$ conformal Killing supervector fields.
Appendix \ref{appendixB} describes another similarity transformation for the AdS supergroup.
Appendix \ref{appendixC} provides an alternative derivation of the (conformal) Killing supervector fields for AdS$^{4|4\N}$.

Our two-component spinor notation and conventions follow  \cite{Buchbinder:1998qv} 
and are similar to those used in \cite{WB}.
In particular, two-component spinor indices are raised and lowered,
\begin{align} \label{raise and lower}
	\psi^{\a} := \ve^{\ab}\psi_{\b}\,, \qquad \psi_{\a} = \ve_{\ab}\psi^{\b}\,; \qquad 
	\bar \phi^{\ad} := \ve^{\ad \bd}\bar \phi_{\bd}\,, \qquad \bar \phi_{\ad} = \ve_{\ad \bd} \bar \phi^{\bd}\,,
\end{align}
 using the spinor metrics
\begin{subequations} \label{epsilon def}
\bea
\ve^{\ab} &=& - \ve^{\b \a} ~, \qquad \ve_{\ab} = - \ve_{\b \a} ~, \qquad \ve^{12}= \ve_{21} =1~; \\
\ve^{\ad \bd} &=& - \ve^{\bd \ad} ~, \qquad \ve_{\ad \bd} = - \ve_{\bd \ad} ~, \qquad \ve^{\dot 1 \dot 2}= \ve_{\dot 2 \dot 1} =1~,
\eea
\end{subequations}
One can convert between vector and spinor indices as follows
\begin{align}
	x_{\a\ad} &= x^{a}(\s_{a})_{\a\ad}\,, \quad x^{\ad\a} = x^{a}(\tilde{\s}_{a})^{\ad\a}
	\quad \Longleftrightarrow \quad x_{a} = -\frac{1}{2}(\tilde{\s}_{a})^{\ad\a}x_{\a\ad}
	= -\frac{1}{2}({\s}_{a})_{\a\ad}x^{\ad\a}\, ,
\end{align}
where the matrices $\s_{a}$ and $\tilde{\s}_{a}$ are given by 
\begin{align}
	\s_{a} = (\id_{2}\,, \vec{\s}) = \big((\s_{a})_{\a\ad}\big)\,, \quad \tilde{\s}_{a} = (\id_{2}\,, -\vec{\s}) =\big((\tilde{\s}_{a})^{\ad\a} \big)\, , \qquad 
	(\tilde{\s}_{a})^{\ad\a} = \ve^{\a\b} \ve^{\ad \bd} (\s_{a})_{\b\bd}\,.
\end{align}


 \section{Three realisations of ${\mathbb F}_1(2) $ } \label{section2}
 
The elements of ${\mathbb F}_1(2)$ are complete flags in ${\mathbb C}^2$. They may be identified with nonsingular $2\times 2$ complex matrices\footnote{We often make use of  antisymmetric $2\times 2$ matrices $ \ve^{ij} $ and $ \ve_{ij} $  normalised as
$\ve^{12}= \ve_{21} =1$.}
\begin{subequations} \label{w-realisation}
\bea
{\bm w} = \big( w_i , v_i) \in \sGL(2,{\mathbb C}) \quad \Longleftrightarrow \quad 
\text{det} \,{\bm w} = v^i w_i \equiv (v, w) \neq 0 ~,
\quad v^i = \ve^{ij}v_j
 \label{w-realisation.a}
\eea
defined modulo equivalence transformations of the form
\bea
{\bm w} \to {\bm w} \tilde{\bm r} \quad \Longleftrightarrow \quad 
v_i \to c v_i ~, \quad w_i \to a w_i + b v_i~, \qquad ac \neq 0~.
 \label{w-realisation.b}
\eea
\end{subequations}
These relations imply that $v_i$ can be interpreted as the homogeneous coordinate for ${\mathbb C}P^1$, while $w_i$ may be made arbitrary modulo the restriction $(v,w) :=v^i w_i \neq 0$. In other words, $w_i$ is a purely `gauge' degree of freedom.

 The same flag manifold, $ {\mathbb F}_1(2) $, can also be realised as 
\bea
{\mathbb F}_1(2) = \sSL(2,{\mathbb C} ) \big/{\mathbb H}_1(2) ~, \qquad
{\mathbb H}_1(2) := \left\{ {\bm r}= \left(
\begin{array}{cc}
 c^{-1}  ~& 0\\
b ~&      c
\end{array}
\right) \in \sSL(2,{\mathbb C}) \right\}~.
\eea
In this realisation, the elements of ${\mathbb F}_1(2)$ are unimodular $2\times 2$ complex matrices
\begin{subequations} \label{v-realisation}
\bea
{\bm v} = \big( v^{-}_i , v^{+}_i) \in \sSL(2,{\mathbb C}) \quad \Longleftrightarrow \quad \text{det} \, {\bm v} = v^{+i} v^{-}_i =1
\eea
defined modulo equivalence transformations of the form
\bea
{\bm v} \to {\bm v} {\bm r} \quad \Longleftrightarrow \quad 
v^{+}_i \to c v^{+}_i ~, \quad v^{-}_i \to c^{-1} v^{-}_i + b v^{+}_i~.
\label{freedom-v}
\eea
\end{subequations}
Here the superscript $\pm$ carried by $v^\pm $ indicates the degree of homogeneity, $v^\pm \to c^{\pm1} v^\pm$, under the scale transformation with parameter $c$.
The above freedom in the  choice of $\bm v$ may be used to choose a representative
\begin{subequations} \label{u-realisation}
\bea
{\bm u} = \big( u^{-}_i , u^{+}_i) \in \sSU(2) \quad  \Longleftrightarrow \quad u^{+i } = \overline{u^-_i}~, \quad 
u^{+i} u^{-}_i =1~.
\eea
There still remain residual equivalence transformations \eqref{freedom-v} of the form 
\bea
u^\pm_i \to \re^{\pm \ri \vf} u^\pm_i~, \qquad \vf \in {\mathbb R}~.
\eea 
\end{subequations}
This leads to the third realisation of the flag manifold $ {\mathbb F}_1(2) $, 
\bea
{\mathbb F}_1(2)  = \sSU(2)\big/\sU(1) \simeq S^2~.
\eea

Both realisations \eqref{w-realisation} and \eqref{v-realisation} are useful when a non-unitary group, 
say $\sGL(2,{\mathbb C})$, acts on ${\mathbb F}_1(2)$. It suffices to consider the action of $\sSL(2,{\mathbb C})$ on ${\mathbb F}_1(2)$, and in this case it is natural to use the realisation 
 \eqref{v-realisation}  of ${\mathbb F}_1(2)$. 
Let $g\approx {\mathbbm 1} + \L \in \sSL(2,{\mathbb C} )$ be an infinitesimal group transformation, 
${\rm tr} \, \L = 0$. Then 
\bea
g {\bm v} \approx {\bm v} + \L{\bm v}~, 
\qquad \L {\bm v} = \big( \L_i{}^j v^-_j ,\, \L_i{}^j v^+_j\big)~.
\eea 
This transformation can be accompanied by an infinitesimal equivalence transformation
\eqref{freedom-v}, 
$ g {\bm v} \sim g {\bm v}{\bm r}$, such that $v^-_i$ remains unchanged. We end up with 
\bea \label{delta v}
\d v^-_i =0~, \qquad \d v^+_i = - \L^{++}(v) v^-_i ~, \quad 
\L^{++}(v) = \L^{jk} v^+_j v^+_k~.
\eea
This transformation law implies that $\sSL(2, {\mathbb C})$ acts on ${\mathbb F}_1(2)$ by holomorphic transformations.

Realisation \eqref{w-realisation} is suitable to understand how the internal space  $T{\mathbb C}P^1$ used in 
 \cite{ButterHSS} originates. We introduce symmetric $2\times 2$ matrices 
\bea
(\S^I)_{i j}=( \ri {\mathbbm 1}, -\s_1, -\s_3) =(\S^I)_{ji}
~,\quad (\S^I)^{i j}:= \ve^{ik} \ve^{jl} (\S^I)_{kl}=
( \ri {\mathbbm 1}, \s_1, \s_3) ~,
\eea
with $I=1,2,3$ and $i,j =12$. 
Their  properties are
 \bea
\sum_{I=1}^3(\S^I)_{ij}(\S^I)_{kl}= 2\ve_{i(k}\ve_{l)j}~, \qquad
(\S^I)_{ij}(\S^J)^{ij}=-2\d^{IJ}~.
\eea
Next we introduce a complex 3-vector 
\begin{subequations}
\bea
\vec{Z} =(Z^I)\in {\mathbb C}^3~, \qquad  Z^I = \frac{1}{(v,w)} v^i (\S^I)_{ij} w^j ~, \label{tangent-a}
\eea
 with the property 
 \bea
 \vec{Z} \cdot \vec{Z}=1~.\label{tangent-b}
 \eea
 \end{subequations}
 The complex hypersurface in  ${\mathbb C}^3 $ defined by eq. \eqref{tangent-b} 
 provides a global realisation for 
   $T{\mathbb C}P^1$. Indeed, if we introduce the real and imaginary parts of $\vec{Z}$, $\vec{Z} = \vec{X} +\ri \vec{Y}$, 
   then the constraint\eqref{tangent-b} can be recast in the form:
   \bea
   \vec{R}\cdot \vec{R} = 1, \qquad  \vec{R} \cdot \vec{Y}=0~,\qquad \vec{R} :=\frac{\vec{X} }{\sqrt{1 + \vec{Y}\cdot \vec{Y}}}~.
 \eea
 The expression for $Z^I$ in terms of $v^i$ and $w^i$, eq.  \eqref{tangent-a}, 
  is invariant under the scale transformations 
  \eqref{w-realisation.b} described by the parameters $a$ and $c$. However $Z^I$ is not invariant under the $b$-transformation \eqref{w-realisation.b}
 
 Of course, the realisation \eqref{v-realisation} is also suitable to describe  the internal space  $T{\mathbb C}P^1$. Here the expression \eqref{tangent-a} for $Z^I$ turns into
\bea
Z^I =  v^+_i (\S^I)^{ij} v^-_j ~.
\eea
This complex three-vector is invariant under the scale $c$-transformation \eqref{freedom-v}.
However, $Z^I$  is not invariant under the $b$-transformation \eqref{freedom-v}.
 
Associated with the three realisations of ${\mathbb F}_1(2)$  considered above are three types of fields. In the case of \eqref{w-realisation}, it is natural to deal with functions $\f^{(p,q)} (v, w)$ that are homogeneous in $v_i$ and, independently, in $w_i$, 
\bea
\f^{(p,q)} (cv, a w) = c^p a^{-q} \f^{(p,q)} (v, w) ~.
\eea
 By replacing $\f^{(p,q)} (v, w) \to (v,w)^q \f^{(p,q)} (v, w) $ we can always make $q=0$, 
 and thus it suffices to work with functions $\F^{(n)} (v, w) \equiv \f^{(n,0)} (v, w)$,
\bea
\F^{(n)} (c v, aw) = c^n \F^{(n)} (v, w) ~.
\label{1.9}
\eea  
Invariance under the $b$-transformations in \eqref{w-realisation.b} will be imposed on an action functional.

Realisation \eqref{v-realisation} is obtained from \eqref{w-realisation} by introducing the variables 
\bea
v^+_i := v_i ~, \qquad v^-_i := (v,w)^{-1} w_i~.
\eea
Then the function $ \F^{(n)} (v, w) $, eq. \eqref{1.9}, turns into $ \F^{(n)} (v^+, v^-)$ such that 
\bea
 \F^{(n)} (c v^+, c^{-1} v^-) = c^n \F^{(n)} (v^+, v^-)~, \qquad 
  c \in {\mathbb C}^* \equiv{\mathbb C} - \{ 0\}~.
\label{1.11}
\eea

Finally, in the case of the harmonic realisation \eqref{u-realisation} one deals with functions over 
$\sSU(2)$, $\J^{(n)} (u^+, u^-)$, of $\sU(1)$ charge $n \in {\mathbb Z}$, with the defining property 
\bea
\J^{(n)}( \re^{\pm\ri \vf} u^\pm) = \re^{ \ri n \vf} \J^{(n)}(u^\pm)~, \qquad \vf \in {\mathbb R} ~.
\eea
 In accordance with Schur's lemma, an arbitrary function $\J(u^\pm)$ over $\sSU(2)$ is a linear combination of functions of definite charge, 
 \bea
 \J(u^\pm) = \sum_{n\in \mathbb Z} \J^{(n)} (u^\pm)~.
\eea
Any function $\J^{(n)} (u^\pm)$ over $\sSU(2)$ proves to be  represented by a convergent Fourier series of the form (see, e.g., \cite{Zhel,GIKOS,GIOS})
\bea
 \J^{(n)}(u^\pm)=\sum_{k=0}^{+\infty}\J^{(i_1\cdots i_{n+k}j_1\cdots j_k)} \,
 u^+_{i_1}\cdots u^+_{i_{k+n}}u^-_{j_1}\cdots u^-_{j_k}~, 
 \label{harmonic_function}
\eea
where the charge is assumed to be non-negative, $n\geq 0$. Analogous representation holds in the $n<0$ case.

As a generalisation of  \eqref{harmonic_function}, 
we formally represent 
 a holomorphic function $ \F^{(n)} (v^\pm) $ satisfying the homogeneity condition  \eqref{1.11} 
as
\begin{subequations}
\bea
 \F^{(n)}(v^\pm) &=& \sum_{k=0}^{+\infty}\F^{(i_1\cdots i_{n+k}j_1\cdots j_k)} \,
 v^+_{i_1}\cdots v^+_{i_{k+n}} v^-_{j_1}\cdots v^-_{j_k}~,  \qquad n\geq 0~,
  \label{holomorphic_function}
 \eea
 where the variables $v^\pm_i$ are related to $u^\pm_i$ as
 \bea
 v^+_i &=& c u^+_i~, \qquad v^-_i = c^{-1} u^-_i +b^{--} u^+_i~, \qquad c \in {\mathbb C}^*
 \label{v-var}
\eea
\end{subequations}
for arbitrary $b^{--} \in \mathbb C$. One may think of $ \F^{(n)}(v^\pm) $ to be an analytic continuation of \eqref{harmonic_function}, assuming that the series in \eqref{holomorphic_function} 
is convergent when $b^{--}$ in \eqref{v-var} is equal to zero. 


 \section{Minkowski flag superspace $ {\mathbb M}^{4|8} \times {\mathbb F}_1(2) $}\label{section3} 

In this section we argue that $ {\mathbb M}^{4|8} \times {\mathbb F}_1(2) $ is suitable to describe off-shell $\cN=2$ supersymmetric theories for all realisations of ${\mathbb F}_1(2) $ discussed earlier.

\subsection{Harmonic superspace approach: the $u^\pm$ realisation}

Within the 4D $\cN=2$ harmonic superspace approach one works with analytic superfields $\cQ^{(n)} (z, u^\pm)$ that are defined on 
${\mathbb R}^{4|8} \times S^2$ and obey the Grassmann analyticity constraints 
\bea
D^+_\a \cQ^{(n)} =0~, \quad \bar D^+_\ad \cQ^{(n)} =0~, \qquad 
D^\pm_\a := u^\pm_i D^i_\a ~, \quad \bar D^\pm_\ad := u^\pm_i \bar D^i_\ad ~.
\eea
With respect to the harmonic variables $u^\pm_i$,  $\cQ^{(n)} (z, u^\pm)$ is a smooth function on $\sSU(2)$ of $\sU(1) $ charge $n$, 
\begin{subequations} \label{Qn-u}
\bea
\cQ^{(n)}( z,\re^{\pm\ri \vf} u^\pm) &=& \re^{ \ri n \vf} \cQ^{(n)}(z,u^\pm)~, \qquad \vf \in {\mathbb R}~, \\
\cQ^{(n)}(z, u)&=&\sum_{k=0}^{+\infty} \cQ^{(i_1\cdots i_{k+n}j_1\cdots j_k)}
(z)\,u^+_{i_1}\cdots u^+_{i_{k+n}}u^-_{j_1}\cdots u^-_{j_k}~, 
\eea
\end{subequations}
where the charge is assumed to be non-negative, $n\geq 0$.
The harmonic superspace action is 
\bea
S \big[\cL^{(+4)} \big]=
\int \rd^4x \int \rd u\,(D^-)^4 \cL^{(+4)}(x, \q, \bar \q, u^\pm)\Big|_{\q=\bar \q=0}~, 
\qquad (D^-)^4 =\frac{1}{16} (D^-)^2 ({\bar D}^-)^2~.
\label{HSSaction}
\eea
The integral over $\sSU(2)$ is defined in accordance with \cite{GIKOS}
\bea
\int \rd u \, \cQ^{(n)} (u^\pm) =  \d_{n, 0} \cQ~.
\label{u-integral}
\eea
Here $Q$ is the harmonic-independent coefficient in the Fourier series for $Q^{(0)}(u^\pm)$, 
\bea
\cQ^{(0)}( u^\pm)&=& \cQ+ \sum_{k=1}^{+\infty}\cQ^{(i_1\cdots i_{k}j_1\cdots j_k)}
\,u^+_{i_1}\cdots u^+_{i_{k}}u^-_{j_1}\cdots u^-_{j_k}~.
\eea

The action \eqref{HSSaction} is known to be $\cN=2$ supersymmetric, see the next subsection for the proof.
 

\subsection{The $v^\pm$ realisation}

Now we analytically continue the superfield \eqref{Qn-u} to the $v$-variables \eqref{v-var}, 
 \begin{subequations} \label{Qn-v}
\bea
\cQ^{(n)}( z, c v^+, c^{-1} v^-) &=& c^n \cQ^{(n)}(z,v^\pm)~, \qquad c \in {\mathbb C}^* \equiv{\mathbb C} - \{ 0\}\\
\cQ^{(n)}(z, v)&=& \cQ(z) +\sum_{k=1}^{+\infty}\cQ^{(i_1\cdots i_{k+n}j_1\cdots j_k)}
(z)\,v^+_{i_1}\cdots v^+_{i_{k+n}}v^-_{j_1}\cdots v^-_{j_k}~, 
\eea
\end{subequations}
and the integer $n$ is said to be the weight of $\cQ^{(n)}$.
This superfield is still Grassmann analytic, 
\bea
D^+_\a \cQ^{(n)} (z,v)=0~, \quad \bar D^+_\ad \cQ^{(n)} (z,v)=0~, \qquad 
D^\pm_\a := v^\pm_i D^i_\a ~, \quad \bar D^\pm_\ad := v^\pm_i \bar D^i_\ad ~.
\label{1.17}
\eea
We can formally extend the algebraic definition of the integral \eqref{u-integral} to the variables $v$, 
 \bea
\int \rd v \, \cQ^{(n)} (v^\pm) :=  \d_{n, 0} \cQ~.
\label{v-integral}
\eea
Finally, we define the flag-superspace action 
\bea
S \big[\cL^{(+4)} \big]=
\int \rd^4x \int \rd v\,(D^-)^4 \cL^{(+4)}(x, \q, \bar \q, v^\pm)\Big|_{\q=\bar \q=0}~,
\label{FlagSSaction}
\eea
where the spinor covariant derivatives $D^-_\a$ and $\bar D^-_\ad$ are defined  in \eqref{1.17}.

 The integrand in \eqref{FlagSSaction}  is obviously invariant under arbitrary rescalings 
$v^{+}_i \to c v^{+}_i$ and $v^{-}_i \to c^{-1} v^{-}_i $. Let us give a small disturbance to the variables $v^-$ 
\bea
\d v^-_i = b^{--} v^+_i ~, 
\label{1.20}
\eea
while keeping $v^+$ fixed.\footnote{Notation $b^{--}$ in \eqref{1.20} indicates that any rescaling $v^+_i \to c v^+_i$ and $\d v^-_i \to c^{-1} \d v^-_i $ results in $b^{--} \to c^{-2} b^{--}$.}
The Lagrangian changes as 
\bea
\d \cL^{(+4)} = b^{--} D^{++}\cL^{(+4)} =  D^{++}\Big(b^{--} \cL^{(+4)} \Big)
~, \qquad D^{++}=v^{+}_i \,\frac{\partial}{\partial v^{- }_i}~.
\eea
Of special importance is the fact that applying the operator $D^{++}$ to any Grassmann analytic superfield $Q^{(n)}$ results in a Grassmann analytic one, 
\bea
\big[D^{++} , D^+_\a \big] =0 ~, \qquad \big[ D^{++}, \bar D^+_\ad \big] =0~.
\eea
It holds that 
\bea
D^+_\a  \X^{++}=0~, \qquad \bar D^+_\ad \X^{++} =0 \quad \implies \quad S\big[D^{++} \X^{++} \big] = 0~.
\eea
We conclude that the action \eqref{FlagSSaction} is invariant under the transformations \eqref{freedom-v}.
Therefore the model is defined on the flag superspace $ {\mathbb M}^{4|8} \times {\mathbb F}_1(2) $,
although the integrand in \eqref{FlagSSaction} is a composite superfield on  $ {\mathbb M}^{4|8} \times T{\mathbb C}P^1 $.

 The action \eqref{FlagSSaction} is supersymmetric. It is actually superconformal 
 provided the Lagrangian $\cL^{(+4)}$ transforms as a primary dimension-2 superfield\footnote{See \cite{Kuzenko:2006mv} for the five-dimensional counterpart of the transformation law \eqref{harmultQ}.}
 \bea
\d_\x \cL^{(4)} = 
\Big(  \x -  \L^{++}[\x] D^{--} \Big)  \cL^{(4)} 
+2 \S [\x] \cL^{(4)} ~,  \qquad D^{--}=v^{-}_i \,\frac{\partial}{\partial v^{+ }_i}~.
\label{harmultQ}
\eea
 Here $\x = \x^A (z) D_A $ is an arbitrary $\cN=2$ conformal Killing supervector field 
 (see Appendix \ref{appendixA}), and $ \L^{++}[\x] $ and $\S[\x]$ are its descendants,
\begin{subequations}\label{W2t3}
\bea
\L^{++} [\x]&:=& v^+_i v^+_j \L^{ij} [\x] ~, 
\\
 \S[\x] &:=& v^+_i v^-_j \L^{ij} [\x]  + \hf (\s[\x]+\bar \s[\x])  ~,
\eea
\end{subequations} 
The descendant  $\L^{ij} [\x]$ and $\s[\x]$ of $\x $ are defined in Appendix \ref{appendixA}.
 The important property of the building blocks \eqref{W2t3}, which appear in  \eqref{harmultQ}, is their 
 Grassmann analyticity
\begin{subequations}
 \bea
 D^+_\a  \L^{++}[\x]=0~, &\qquad &\bar D^+_\ad \L^{++} [\x]=0 ~, \\
 D^+_\a  \S[\x]=0~, &\qquad &\bar D^+_\ad \S [\x]=0 ~.
 \eea
 \end{subequations}

To massage the variation \eqref{harmultQ}, we point out  the identity 
\bea
\x = {\overline \x} = \x^a (z) \pa_a 
- \big( \x^{+\a} D^-_\a + {\bar \x}^{+ \ad} {\bar D}^-_{ \ad} \big)
+ \big( \x^{-\a} D^+_\a + {\bar \x}^{- \ad} {\bar D}^+_{ \ad} \big)~,~~~
\label{xi-2}
\eea
with $\x^{\pm \a} =\x^{\a i} \,v^\pm_i $ and 
${\bar \x}^{+ \ad} ={\bar \x}^{ \ad  i} \,v^\pm_i$.
 Then, making use of the properties of $\x$, \eqref{harmultQ}
  may be brought to the form:
 \bea
\d \cL^{(4)} = \pa_a \big(\x^a \cL^{(4)}\big)
+
D^-_\a\big( \x^{+\a}L^{(4)} \big)  + 
{\bar D}^-_{\ad}  \big( {\bar \x}^{+ \ad} \cL^{(4)}\big)
-D^{--}\big(\L^{++}[\x] \cL^{(4)}\big)~,
\label{harmultQ2}
\eea
see \cite{Kuzenko:2006mv, Kuzenko:2007qy} for similar derivations. 
Here the first three terms on the right do not contribute to the variation of the action, 
\bea
\d_\x S \big[\cL^{(+4)} \big]=
\int \rd^4x \int \rd v\,(D^-)^4 \d_\x\cL^{(+4)}(x, \q, \bar \q, v^\pm)\Big|_{\q=\bar \q=0}~.
\label{FlagSSaction2}
\eea
The last term in \eqref{harmultQ2} also does not contribute to the variation of the action since 
\bea
\int \rd v \,D^{--} \cQ^{++} (v^\pm ) =0~.
\eea
  
Analysing the transformation law \eqref{harmultQ}, one observes that it includes a transformation of the complex harmonics $v^\pm_i$ of the form \eqref{delta v}.
  

\subsection{Projective superspace approach: the $ \big( w , v)$ realisation}

In this approach, off-shell supermultiplets are described in terms of weight-$n$ Grassmann analytic
superfields $Q^{(n)}(z,v)$,
\bea
 {D}^{(1)} _{ \a} Q^{(n)} &=& {\bar D}^{(1)}_{\ad } Q^{(n)} =0~, \quad
  Q^{(n)} (z,c\, v )= c^{n}  \, Q^{(n)}(z,v)~,
\quad c\in {\mathbb C}^* 
\label{5.21}
 \eea
which are independent of $w$,
\bea
\frac{\pa}{\pa w_i} \, Q^{(n)} =0~.
\eea
In other words, $Q^{(n)}(z,v)$ is a holomorphic function on an domain of ${\mathbb C}P^1$, with $v_i$ being the homogeneous coordinates for ${\mathbb C}P^1$.

The projective-superspace action principle is\footnote{In the super-Poincar\'e case,
this action was introduced in \cite{KLR}.
It was re-formulated in a 
manifestly  projective-invariant form in \cite{S}. The superconformal case was studied in \cite{Kuzenko:2007qy}.} 
\bea
S:= - \frac{1}{2\p} \oint_{\g}  { v_i {\rm d} v^i }
\int {\rm d}^4x \, \D^{(-4)} L^{(2)} (z,v) \Big|_{\q={\bar \q} =0}~. 
\label{PAP}
\eea
Here $\g$ denotes a closed contour  in  ${\mathbb C}P^1$, $v^i(t)$,
parametrized by an evolution parameter $t$.
The action makes use of the following fourth-order differential operator:
\bea
\D^{(-4)} := \frac{1}{16} \nabla^\a \nabla_\a {\bar \nabla}_{ \bd}  {\bar \nabla}^{\bd} ~, \quad
\nabla_\a := \frac{ 1}{ (v,w)} 
{ w_i} D^i_\a 
~, \quad 
{\bar \nabla}_{\bd} := \frac{1 }{(v,w)} 
w_i{\bar D}^i_{\bd} ~,~~
\eea
where $(v,w):= v^i w_i$.
Here $w_i$ is a {fixed} isotwistor chosen to be arbitrary modulo 
the condition $(v,w) \neq 0$ along the integration contour. The action proves to be independent of $w_i$, 
see \cite{Kuzenko:2007qy} for the proof.  Thus the action is invariant under 
arbitrary transformations \eqref{freedom-v}.


\section{New realisation of the AdS supergroup} \label{section4}

In this section we introduce a new realisation for the connected component
$\mathsf{OSp}_0({\cal N}|4; {\mathbb R})$ of the AdS isometry supergroup 
 as $\mathsf{SU}(2,2 |{\cal N}) \bigcap \mathsf{OSp} ({\cal N}| 4; {\mathbb C})$. It will be used in Section \ref{section5}.

\subsection{The superconformal group and supertwistors} \label{section4.1}

The $\N$-extended superconformal group in four dimensions is $\sSU(2,2|\N)$.\footnote{The case $\cN=4$ is somewhat special, but the corresponding details  will not be discussed here.} 
By definition, it consists of all supermatrices 
\begin{align}
	\hat{g} = \left(\begin{array}{c||c}
A & B \\
\hline \hline
C & D 
\end{array}\right)	\,, \qquad \hat{g} \in \sSL(4|\N;{\mathbb C})
\end{align}
satisfying the master equation 
\begin{align}
	\hat{g}^{\dag} \O \hat{g} = \O\,, 
	\qquad \O = 
	\left(\begin{array}{c|c||c}
	0 & \id_{2} & 0 \\
	\hline 
	\id_{2} & 0 & 0 \\
	\hline \hline 
	0 & 0 & -\id_{\N}
	\end{array}\right)\,.
\end{align}

In accordance with \cite{Ferber}, a supertwistor $T$ is a column vector 
\begin{align}
T = (T_{A}) = \left(\begin{array}{c}
T_{\ah}\\ \hline \hline
T_{i}
\end{array}\right)\,, \qquad \ah = 1\,,2\,,3\,,4\,, \qquad i = 1\,, \ldots \,, \N\,.
\end{align}
In the case of {\it even} supertwistors, $ T_\ah$ is bosonic
and $T_i$ is fermionic.
In the case of {\it odd} supertwistors, $T_\hal$ is fermionic while $T_i$ is bosonic.
The even and odd supertwistors are called pure.
We introduce the parity function $\e ( T )$ defined as:
$\e ( T ) = 0$ if $ T$ is even, and $\e ( T ) =1$ if $T$ is odd.
If we define
\begin{align}
	\e_A= \bigg\{\begin{array}{ccc}
		0 ~,& ~ & A = \ah \\
		1 ~,& ~ & A = i
	\end{array}
\end{align}
then the components $T_A$ of a pure supertwistor
 have the following  Grassmann parities
\bea
\e ( T_A) = \e ( T ) + \e_A \quad (\mbox{mod 2})~.
\eea
The space of even supertwistors is naturally identified with
${\mathbb C}^{4|\cN}$,
while the space of odd supertwistors may be identified with
${\mathbb C}^{\cN |4}$.
The supergroup $\sSU(2,2|\N)$ acts on the space of {even} supertwistors and on the space of {odd} supertwistors,
\begin{align}
T \rightarrow \hat{g}T \quad \implies \quad  T^{\dag}\O \rightarrow T^{\dag}\O\, \hat{g}^{-1}\,. 
\end{align} 
It holds that $\e(\hat{g}T ) = \e(T)$.
The supertwistor space is equipped with the $\sSU(2,2|\N)$-invariant   inner product 
\begin{align} \label{scf ip}
\braket{T}{S} = T^{\dag}\O S \,.
\end{align}
%


\subsection{The AdS supergroup}

In this paper, the connected component $\sOSp_0(\N|4;\mathbb{R})$
of  $\sOSp(\N|4;\mathbb{R})$
will be  identified with the 
$\N$-extended AdS supergroup
in four dimensions. 
We recall that the supergroup $\sOSp(\N|4;\mathbb{C})$
consists of those supermatrices 
\begin{align}
	f =(f_A{}^B )= \left(\begin{array}{c||c}
A & B \\
\hline \hline
C & D 
\end{array}\right)	 \in \sGL(4|\N;{\mathbb C})
\end{align}
which satisfy the master equation
\bsubeq \label{old ads group}
\begin{align} 
	f^{\sT} \mathbb{J} f &= \mathbb{J}\,,
	\end{align} 
where $f^{\sT}$ denotes the super-transpose of $f$, 
 \begin{align}
(f^{\rm sT})^A{}_B := (-1)^{\e_A \e_B + \e_B} f_B{}^A \quad \Longleftrightarrow \quad
	f^{\sT} = \left(\begin{array}{c||c}
A^{\T} & -C^{\T} \\
\hline \hline
B^{\T} & D^{\T} 
\end{array}\right)\,,
\end{align} 
and the symplectic supermatrix $\mathbb J$ is given by 
\begin{align}
	\qquad \mathbb{J} = 
	\left(\begin{array}{c|c||c}
	0 & \id_{2} & 0 \\
	\hline 
	-\id_{2} & 0 & 0 \\
	\hline \hline 
	0 & 0 & \ri\id_{\N}
	\end{array}\right)\,.
\end{align}
The elements of  $\sOSp(\N|4;\mathbb{R}) \subset \sOSp(\N|4;\mathbb{C})$ satisfy the reality condition 
\begin{align} 
	f^{\dag} &= f^{\sT}\,.
\end{align}
\esubeq

The supergroup $\sOSp(\N|4;\mathbb{C})$ naturally acts on the supertwistor space.\footnote{The supertwistor space is   defined as in the previous subsection. However, in this subsection our attention is restricted to the action of $\sOSp(\N|4;\mathbb{C})$
or of its subgroup of  $\sOSp(\N|4;\mathbb{R})$ on the supertwistor space.}
This action is characterised by the $\sOSp(\N|4;\mathbb{C})$-invariant inner product
\bsubeq
\begin{align}
	\braket{T}{S}_{\mathbb{J}} &:= T^{\sT}\mathbb{J} S\,,
\end{align}
\esubeq
where the supertranspose  $T^{\sT}$ of $T$ is defined as 
\begin{align}
T^{\sT} := \left(T_{\ah}\,, -(-1)^{\e(T)}T_{i}\right) = (T_{A}(-1)^{\e(T)\e_{A} + \e_{A}})\,.
\end{align}

Now, let us restrict our attention to the action of  $\sOSp(\N|4;\mathbb{R})$ on the supertwistor space.
Then there exists the involution $*$ defined as 
\begin{align} \label{original conjugate}
T \rightarrow *T\,, \qquad (*T)_{A} = (-1)^{\e(T)\e_{A} + \e_{A}}\overline{T_{A}}\,, 
\end{align}
where $\overline{T_{A}}$ denotes the complex conjugate of $T_{A}\,.$
Its key properties are 
\bsubeq
\begin{align}
	* (* T) &= T\,, \\
	f(* T) &= *(f T)\,, \qquad \forall f \in  \sOSp(\N|4;\mathbb{R})~.
\end{align}
\esubeq
A supertwistor is said to be real if it satisfies the reality condition
\begin{align}
*T = T \quad \Longleftrightarrow \quad T^{\dag} = T^{\sT}\,.
\end{align}


\subsection{New realisation of the AdS supergroup}\label{section new ads group}

For our purposes it is useful to work with an alternative realisation of the AdS supergroup, as a subgroup of the superconformal group. 
Let 
$\sOSp_0(\N|4;\mathbb{R})_{\mathfrak U}$
be 
the subgroup of $\sSU(2,2|\N)$
consisting of those  supermatrices $g \in  \sSL(4|\N;{\mathbb C})$
which are singled out by the conditions
\bsubeq \label{new group con}
\begin{align}
g^{\dag} \O g &= \O \,,
\\ 
g^{\sT} \mathfrak{J} g &= \mathfrak{J}\,,
\end{align}
\esubeq
where $\mathfrak{J}$ denotes follows symplectic supermatrix
\begin{align}
	\mathfrak{J} = \left(\begin{array}{c|c||c}
		\ve & ~0~ & 0 \\
		\hline 
		~0~ & -\ve & 0 \\
		\hline \hline 
		0 & 0 & \ri \id_{\N}
	\end{array}\right)\,, 
	\qquad \ve = 
	 \left(\begin{array}{cc}
		~0~ & 1  \\
		-1 & ~0~ 
	\end{array}\right)\,. 
\end{align}

The supergroup $\sOSp_0(\N|4;\mathbb{R})_{\mathfrak U}$ proves to be isomorphic to $\sOSp_0(\N|4;\mathbb{R})\,.$ 
The proof is based on the following supermatrix correspondence 
\begin{align} \label{g def}
	f \rightarrow g = {\mathfrak U}^{-1} f \mathfrak{U}\,, \qquad \forall f \in \sOSp_0(\N|4;\mathbb{R})\,.
\end{align}
Here the supermatrix $\mathfrak U$ is defined as 
\begin{align} \label{u def}
	\mathfrak{U} &= \left(\begin{array}{c||c}
	\mathfrak{m} & ~0~ \\
	\hline \hline 
	~0~ &  \id_{\N}
\end{array}\right)\,, \qquad \mathfrak{m} = \frac{1}{2}\left(\begin{array}{c|c}
	\a \id_{2} + \bar{\a}\ve  &  \bar{\a} \id_{2} + \a \ve \\ \hline
	-\a \id_{2} + \bar{\a}\ve & \bar{\a}\id_{2} - \a \ve
\end{array}\right)\,, \qquad \a = \re^{\ri \p/4}
= \frac{1+\ri}{\sqrt{2}}\,. 
\end{align}
It obeys the useful relations 
\bsubeq \label{useful props}
\begin{align}
	\mathfrak{U}^{\dag} &= \mathfrak{U}^{-1}\,, 
	\\
	\mathfrak{U}^\dagger \mathbb{J}  \mathfrak{U}  &= -\ri \O\,, 
	\\
	(\mathfrak{U})^{\sT}\mathbb{J} \mathfrak{U} &=  \mathfrak{J}\,.
\end{align}
\esubeq
It can be constructed making use of the alternative realisations for $\sOSp_0(\N|4;\mathbb{R})$ and $\sSU(2,2|\N)$ provided in \cite{Kuzenko:2006mv} and \cite{Buchbinder:2015qsa}. 
Specifically, 
\begin{align}
\frak{U}^{-1} = M \S \mathbb{U}\,,
\end{align}
with 
\bsubeq
\begin{align}
M &= \frac{1}{\sqrt{2}}\left(\begin{array}{c|c||c}
\id_{2} & -\ve & 0 \\ 
\hline 
-\ve & \id_{2} & 0 \\ \hline \hline 
~0~ & ~0~ & \sqrt{2}\id_{\N}
\end{array}\right)\,, 
\\
\S &= 
\frac{1}{\sqrt{2}}\left(\begin{array}{c|c||c}
\id_{2} & -\id_{2} & 0 \\ 
\hline 
\id_{2} & \id_{2} & 0 \\ \hline \hline 
~0~ & ~0~ & \sqrt{2}\id_{\N}
\end{array}\right)\,, 
\\
\mathbb{U} &= 
\frac{1}{\sqrt{2}}\left(\begin{array}{c|c||c}
\id_{2} & \ri\id_{2} & 0 \\ 
\hline 
\ri\id_{2} & \id_{2} & 0 \\ \hline \hline 
~0~ & ~0~ & \sqrt{2}\id_{\N}
\end{array}\right)\,. 
\end{align}
\esubeq
The relations \eqref{useful props} can be proven with the aid of the following properties 
\bsubeq
\begin{align}
\mathbb{U} \mathbb{J} \mathbb{U}^{\dag} &= -\ri \mathbb{I}\,, \qquad (\mathbb{U}^{-1})^{\sT}\mathbb{J} \mathbb{U}^{-1} = \mathbb{J}\,, 
\\
\S \mathbb{I} \S^{\dag} &= \O\,, \qquad ~~(\S^{-1})^{\sT}\mathbb{J} \S^{-1} = \mathbb{J}\,, \\
M \O M^{\dag} &= \O\,, \qquad  (M^{-1})^{\sT} \mathbb{J} M^{-1} = \frak{J}\,,
\end{align}
\esubeq
where $\mathbb{I}$ is defined as 
\begin{align}
\mathbb{I} = \left(\begin{array}{c|c||c}
		\id_{2} & ~0~ & 0 \\
		\hline 
		~0~ & -\id_{2} & 0 \\
		\hline \hline 
		0 & 0 & - \id_{\N}
	\end{array}\right)\,. 
\end{align}
Further, the matrices $M$, $\S$ and $\mathbb{U}$ are unitary,
\begin{align}
M^{-1} = M^{\dag}\,, \qquad \S^{-1} = \S^{\dag}\,, \qquad \mathbb{U}^{-1} = \mathbb{U}^{\dag}\,. 
\end{align}
These properties imply that the supermatrix $g$ defined by eq. \eqref{g def} obeys the conditions \eqref{new group con}, and hence $g \in \sOSp_0(\N|4;\mathbb{R})_{\frak{U}}\,.$
In Appendix \ref{appendixB} we introduce another supermatrix that relates the realisations 
$ \sOSp_0(\N|4;\mathbb{R})$ and $ \sOSp_0(\N|4;\mathbb{R})_{\frak{U}}\,.$

Associated with $\sOSp_0(\N|4;\mathbb{R})_{\mathfrak U}$ are two invariant inner products 
\bsubeq
\begin{align}
\braket{T}{S}_{\O} &:= T^{\dag} \O S\,, 
\\
\braket{T}{S}_{\mathfrak{J}} &:= T^{\sT}\mathfrak{J}S\,,
\end{align}
\esubeq
for arbitrary pure supertwistors $T$ and $S$.

The supergroup elements $g$ satisfy the reality condition 
\begin{align} \label{new ads reality}
	g^{\dag} = \U^{-1}g^{\sT}\U \,, \qquad \U = \frak{U}^{\sT}\frak{U} = \left(\begin{array}{c|c||c}
	~0~ & \ri\ve & ~0~ \\ 
	\hline 
	-\ri\ve & ~0~ & 0 \\ 
	\hline \hline 
	0 & 0 & \id_{\N}
	\end{array}\right)\,.
\end{align}
Then, making use of eq. \eqref{new ads reality}, one can introduce an involution operation $\star$ defined as 
\begin{align}
	T \rightarrow \star T\,, \qquad 	(\star T)_{A} = (-1)^{\e(T)\e_{A} + \e_{A}}\left( \U^{-1}\overline{T} \right)_{A}\,.  
\end{align}
Its key properties are 
\bsubeq
\begin{align}
	\star (\star T) &= T\,, \\
	g(\star T) &= \star(g T)\,. 
\end{align}
\esubeq
In our new realisation of the AdS supergroup, a supertwistor $T$ is said to be real if it satisfies 
\begin{align}
\star T = T\,.
\end{align}
Further, we observe that 
\begin{align}
\star{T}^{\sT} = T^{\dag} \U^{-1}\,,
\end{align}
which, in conjunction with the relations \eqref{useful props}, yields 
\begin{align}
\braket{\star{T}}{S}_{\frak{J}} = - \ri\braket{T}{S}_{\O}\,.
\end{align}

\section{The supertwistor realisations of AdS$^{4|4\N} $}\label{section5}

Before introducing the supertwistor realisation of AdS$^{4|4\N}$ in terms of the supergroup $\sOSp_0(\N|4;\mathbb{R})_{\frak{U}}$, we recall the original construction given in  \cite{Kuzenko:2021vmh} and formulated 
in terms of the supergroup $\sOSp_0(\N|4;\mathbb{R})\,.$ 

\subsection{Original realisation}

Here we will make use of the supergroup $\sOSp_0(\N|4;\mathbb{R})$. 
Let us consider the space of complex even supertwistors, which can be identified with  $\mathbb{C}^{4|\N}$. 
In this space, we consider complex two-planes which are generated by two even supertwistors
\begin{align}
\mathfrak{T} = (T_{A}{}^{\m})\,, \qquad \m = 1\,,2\,,
\end{align}
such that the bodies of $T^{1}$ and $T^{2}$ are linearly independent. 
By construction, the supertwistors $T^{\m}$ are defined modulo the equivalence relation 
\begin{align} \label{original ads equiv}
T^{\m} \sim \tilde{T}^{\m} = T^{\n}R_{\n}{}^{\m}\,, \qquad R = (R_{\n}{}^{\m}) \in \sGL(2\,,\mathbb{C})\,,
\end{align}
as the bases $\{T^{\m}\}$ and $\{\tilde{T}^{\m}\}$ span the same two-plane. 
We restrict our attention to those two-planes which satisfy the constraints 
\bsubeq \label{original ads constraints}
\begin{align}
\ve_{\m\n}\braket{T^{\m}}{T^{\n}}_{\mathbb{J}} & \neq  0\,,
\\
\braket{*T^{\m}}{T^{\n}}_{\mathbb{J}} & = 0\,. 
\end{align}
\esubeq
Here the supertwistor $*T$ denotes the conjugate of $T$ defined by 
\eqref{original conjugate}. 
These conditions are preserved under the action of the supergroup $\sOSp_0(\N|4;\mathbb{R})\,.$
We say that any pair of supertwistors satisfying the constraints \eqref{original ads constraints} constitutes a frame, and the space of frames is denoted $\frak{F}_{\N}$. 

The supergroup $\sOSp_0(\N|4;\mathbb{R})$ acts on the space of frames as 
\begin{align}
T^{\m} \rightarrow fT^{\m}\,, \qquad f \in \sOSp_0(\N|4;\mathbb{R})\,.
\end{align}
This action is naturally extended to the quotient space $\frak{F}_{\N}/ \sim$, where the equivalence relation is given by \eqref{original ads equiv}. 
As shown in \cite{Kuzenko:2021vmh}, AdS$^{4|4\N}$ can be identified with this quotient space
\begin{align}
\text{AdS}^{4|4\N} = \frak{F}_{\N}/\sim \,.
\end{align}
%


\subsection{New realisation}

In this subsection we will show how the AdS superspace AdS$^{4|4\cal N}$ arises  an open domain  of compactified ${\cal N}$-extended Minkowski superspace, $\overline{\mathbb M}^{4|4\cal N}$, the latter being studied in \cite{Kuzenko:2006mv}.

As discussed in \cite{Kuzenko:2006mv}, 
$\overline{\mathbb M}^{4|4\cal N}$ is the space of null two-planes in the space of complex even supertwistors. 
Given such a two-plane, it may be described by two even supertwistors 
\begin{align}
\mathfrak{T} = (T_{A}{}^{\m})\,, \qquad \m = 1\,,2\,,
\end{align}
such that the bodies of $T^{1}$ and $T^{2}$ are linearly independent. 
That the two-planes are null means they satisfy the constraint 
\begin{align} \label{null con}
\braket{T^{\m}}{T^{\n}}_{\O} = 0\,.
\end{align}
By construction, the supertwistors $T^{\m}$ are defined modulo the equivalence relation 
\begin{align} \label{gl equiv}
T^{\m} \sim \tilde{T}^{\m} = T^{\n}R_{\n}{}^{\m}\,, \qquad R = (R_{\n}{}^{\m}) \in \sGL(2\,,\mathbb{C})\,,
\end{align}
as the bases $\{T^{\m}\}$ and $\{\tilde{T}^{\m}\}$ span the same two-plane.
The condition \eqref{null con} is preserved under the action of the superconformal group $\sSU(2,2|\N)$.

Let us restrict our attention to those two-planes which satisfy the additional condition 
\begin{align} \label{symp neq 0}
\braket{T^{\m}}{T^{\n}}_{\mathfrak{J}} \neq 0\,. 
\end{align}
Then, making use of the equivalence relation \eqref{gl equiv}, we can normalise the two-planes such that 
\bsubeq \label{ads st con}
\begin{align} 
\braket{T^{\m}}{T^{\n}}_{\O} &= 0\,. 
\\
\braket{T^{\m}}{T^{\n}}_{\mathfrak{J}} &= \ell \ve^{\m\n}\,, \label{normal con}
\end{align}
\esubeq
for some constant parameter $\ell > 0\,.$ 
The conditions \eqref{ads st con} are preserved under the action of the AdS supergroup $\sOSp_0(\N|4;\mathbb{R})_{\frak{U}}$, and under equivalence transformations of the form 
\begin{align} \label{sl equiv}
T^{\m} \sim \tilde{T}^{\m} = T^{\n}N_{\n}{}^{\m}\,, \qquad N = (N_{\n}{}^{\m}) \in \sSL(2\,,\mathbb{C})\,.
\end{align}
We say that any pair of supertwistors satisfying the relations \eqref{ads st con} constitutes a frame, and the space of frames is denoted $\frak{F}_{\N}\,.$
The supergroup $\sOSp_0(\N|4;\mathbb{R})_{\frak{U}}$ acts on $\frak{F}_{\N}$ by the rule 
\begin{align}
T^{\m} \rightarrow gT^{\m}\,, \qquad g \in \sOSp_0(\N|4;\mathbb{R})_{\frak{U}}\,.
\end{align}
This action is naturally extended to the quotient space $\frak{F}_{\N}/\sim$, which was identified with AdS superspace in \cite{Kuzenko:2021vmh}, 
\begin{align}
\text{AdS}^{4|4\N} = \frak{F}_{\N}/\sim\,. 
\end{align}

\subsection{The north chart of AdS$^{4|4\N}$}

In what follows, we will set $\ell = 1\,.$ 
It is instructive to write the two-plane explicitly as 
\begin{align}
\mathfrak{T} = (T^{\m}) = \left(\begin{array}{c}
F\\ 
G \\ \hline \hline 
\vf
\end{array}\right) \,,
\end{align}
Here, $F$ and $G$ are $2\times 2$ matrices, and $\vf$ is an $\N\times 2$ matrix.  
Then, the conditions \eqref{ads st con} imply the following 
\bsubeq \label{block con}
\begin{align}
F^{\dag}G + G^{\dag}F - \vf^{\dag}\vf &= 0\,, \label{det con}
\\
(\det F - \det G)\ve - \ri\vf^{\T}\vf &= \ve\,. 
\end{align}
\esubeq

Let us define the north chart to consist of those normalised two-planes with $\det F \neq 0\,.$ 
Then, making use of the equivalence relation \eqref{sl equiv}, we can choose 
\begin{align}
F = \l \id_{2}\,,
\end{align}
for some parameter $\l\,.$
We then we find that the two-planes are of the form
\begin{align} \label{nc 2p}
\mathfrak{T}  = \l \left(\begin{array}{c}
\id_{2} \\ 
-\ri \tilde{x}_{+} \\ \hline \hline 
2\q
\end{array}\right)\,, \quad \tilde{x}_{+} =\big( x_{+}^{\ad\a}\big)\, , \quad \q =\big( \q_i{}^\a\big)\,,
\end{align}
where the bosonic $\big( x_{+}^{\ad\a}\big) $ and fermionic $\big( \q_i{}^\a\big) $variables are {\it chiral} and satisfy the condition 
\begin{align}
	\tilde{x}_{+} - \tilde{x}_{-} = 4\ri\q^{\dag}\q\,, \qquad \tilde{x}_{-} = (\tilde{x}_{+})^{\dag}\,.
\end{align}
The solution to the above condition is given by
\begin{align}
\tilde{x}_{\pm} = \tilde{x} \pm 2\ri \q^{\dag}\q\,, \qquad \tilde{x}_{\pm} = x_{\pm}^{a}\tilde{\s}_{a}\,, \qquad \tilde{\s}_{a} = (\id_{2}\,, - \vec{\s})\,, 
\end{align}
where $\vec{\s}$ are the Pauli matrices. 
The parameter $\l$ takes the form 
\begin{align} \label{lambda def}
\l = (1 - x_{+}^{2} + 2\ri\q^{2})^{-\frac{1}{2}}\,, \qquad \q^{2} = \tr (\q^{\T}\q \ve)\,. 
\end{align}
It follows that the north chart is parametrised by the chiral coordinates $x_{+}^{a}$ and $\q_{i}{}^{\a}\,.$

To describe the action of the AdS supergroup on the north chart, it is instructive to begin with an element of the superconformal group, which can be represented as 
\begin{align} \label{scf alg}
	\hat{g} = \re^{L}\,, 
	\qquad L = 
\left(\begin{array}{c|c||c}
	-K_{\a}{}^{\b} - \hf\D\d_{\a}{}^{\b} & \ri b_{\a\bd} & 2\eta_{\a}{}^{j} \\
	\hline
	-\ri a^{\ad\b} & ~\bar{K}^{\ad}{}_{\bd} +\hf \bar{\D}\d^{\ad}{}_{\bd} & 2\bar{\e}^{\ad j} \\
	\hline \hline
	2\e_{i}{}^{\b} & 2\bar{\eta}_{i\bd} & \frac{1}{\N}(\bar{\D}-\D)\d_{i}{}^{j} + \L_{i}{}^{j}
\end{array}
\right)\,, 
\end{align}
with 
\begin{align}
K = (K_{\a}{}^{\b})\,, \qquad \tr K = 0\,, \qquad  \L = (\L_{i}{}^{j})\,, \qquad \L^{\dag} = -\L \,, \qquad \tr \L = 0\,. 
\end{align}
Here, the matrix elements correspond to a Lorentz transformation $(K_{\a}{}^{\b}\,, \bar{K}^{\ad}{}_{\bd})$, Poincar\'e translation $a^{\ad\b}$, special conformal transformation $b_{\a\bd}$, $Q$-supersymmetry $(\e_{i}^{\a}\,,\bar{\e}^{\ad i})$, $S$-supersymmetry $(\eta_{\a}^{i}\,, \bar{\eta}_{i \ad})$, combined chiral and scale transformation $\D$, and $\sSU(\N)$ transformation $\L_{i}{}^{j}\,.$ 

It can be shown (see \cite{Kuzenko:2006mv} for the derivation) that, under infinitesimal superconformal transformations, the coordinates of the north chart transform as 
\bsubeq \label{nc chiraltra}
\bea
\d \tilde{x}_+ &=& \tilde{a}  + \hf (\D +{\bar \D})\, \tilde{x}_+
+{\bar K} \tilde{x}_+ +\tilde{x}_+ K 
- \tilde{x}_+ \,b \,\tilde{x}_+
+4{\rm i}\, {\bar \e} \, \q - 4 \tilde{x}_+ \, \eta \, \q ~,
\\
\d \q &=& \e + \frac{1}{2\cN} \Big( 
(\cN-2) \D + 2 
{\bar \D}\Big)\, \q + \q K 
+ \L \, \q  - \q \, b \,  \tilde{x}_+
-{\rm i}\,{\bar \eta}\, \tilde{x}_+ - 4\,\q \,\eta \, \q~.
\eea
\esubeq
The AdS transformations can be singled out as those superconformal transformations which preserve the AdS condition \eqref{normal con}. 
This requirement proves to impose the following constraints on the parameters in \eqref{nc chiraltra}
\bsubeq \label{ads translation con}
\begin{align}
	b^{a} &= - a^{a}\,, 
	\\
	\eta_{\a}{}^{i} &= \ri\d^{ij}\e_{\a j}\,,
	\\
	\D &= 0\,.
\end{align}
\esubeq
Further, only the antisymmetric component of $\L$ remains
\begin{align} \label{ads so(n) con}
\L = -\L^{\T}\,. 
\end{align}

\subsection{Invariant supermetric on AdS$^{4|4\N}$}

In this section we will elucidate some more details about the supertwistor construction above, and use it to introduce an $\sOSp_0(\N|4;\mathbb{R})_{\frak{U}}$-invariant  supermetric on AdS$^{4|4\N}$. Our analysis is similar to that given in \cite{Kuzenko:2014yia} for the  $2n$-extended supersphere $S^{3|4n}$,

Given a superconformal transformation $\hat{g} \in \sSU(2,2|\N)$ that preserves the condition  \eqref{symp neq 0}
on an open domain of AdS$^{4|4\cal N}$, a two-plane $\mathfrak{T}$ transforms as 
\begin{align}
\mathfrak{T} ~\rightarrow ~\hat{g}\mathfrak{T} \sim \mathfrak{T}' = \hat{g}\mathfrak{T}R(\hat{g}\,,\mathfrak{T})\,, \qquad 
R(\hat{g}\,,\mathfrak{T}) \in \sGL(2\,,\mathbb{C})\,.
\end{align}
Here, the matrix $R(\hat{g}\,,\mathfrak{T})$ serves two purposes: (i) it is used to preserve the parametrisation of $\mathfrak{T}$, 
eq. \eqref{nc 2p}; and (ii) it is used to restore the normalisation condition \eqref{normal con}. 
Indeed, for a generic superconformal transformation, the two-plane $\hat{g}\mathfrak{T}$ does not satisfy \eqref{normal con}. 
However, provided it still satisfies \eqref{symp neq 0}, that is
\begin{align}
\braket{\hat{g}T^\m }{\hat{g}T^\n}_{\frak{J}} \neq 0\,,
\end{align}
one can always make use of the equivalence relation \eqref{gl equiv} to restore the normalisation
\begin{align}
\braket{T'^\m}{T'^\n}_{\frak{J}} = \ell\ve^{\m\n}\,. 
\end{align}

The situation differs slightly for AdS transformations. Given an element of the AdS supergroup $g \in \sOSp_0(\N|4;\mathbb{R})_{\frak{U}}$, a two-plane $\mathfrak{T}$ transforms as 
\begin{align} \label{finite ads trf}
\mathfrak{T}~ \rightarrow~ g\mathfrak{T} \sim \mathfrak{T}' = g\mathfrak{T}N(g\,,\mathfrak{T})\,, 
\qquad N(g\,,\mathfrak{T}) \in \sSL(2\,,\mathbb{C})\,. 
\end{align}
That the matrix $N(g\,,T)$ belongs to $\sSL(2\,,\mathbb{C})$ follows from the fact that the AdS transformations preserve the condition \eqref{normal con}. 
To prove this, let us consider a two-plane belonging to the north chart of AdS$^{4|4\N}$, eq. \eqref{nc 2p}. 
Then, for the AdS transformation $g$, we have 
\begin{align}
g\mathfrak{T} = \l(x_{+}\,,\q) \left(\begin{array}{c}
A(g\,,x_{+}\,,\q) \\
B(g\,,x_{+}\,,\q) \\
\hline \hline
\c(g\,,x_{+}\,,\q)
\end{array}\right)\,.
\end{align}
Here we have explicitly indicated the dependence of $\l$ on the coordinates $x_{+}^{a}$ and $\q_{i}{}^{\a}$. 
Further, the matrices $A\,,B\,,\c$ are coordinate-dependent as well as determined by the specific transformation $g$ under consideration. 
For simplicity, let us assume that $gT$ also belongs to the north chart of AdS$^{4|4\N}$, that is $\det A \neq 0\,.$ 
We then introduce the unimodular matrix 
\begin{align}
N  = A^{-1} \det A^{\frac{1}{2}}  ~\implies ~\det N = 1\,. 
\end{align}
Making use of the equivalence relation \eqref{sl equiv}, we have 
\begin{align}
g\mathfrak{T} \sim g\mathfrak{T} N = \l(x_{+}\,,\q) \det A^{\frac{1}{2}} \left(\begin{array}{c}
\id_{2} \\ 
B A^{-1} \\
\hline \hline 
\c A^{-1}
\end{array}\right) 
\equiv 
\g(x'_{+}\,,\q') \left(\begin{array}{c}
\id_{2}
\\
-\ri \tilde{x}_{+}'
\\
\hline \hline 
2\q'
\end{array}\right)\,. 
\end{align}
Finally, the symplectic condition \eqref{normal con} implies 
\begin{align}
\g(x_{+}'\,, \q') = (1 - x_{+}'^{2} + 2\ri\q'^{2})^{-\frac{1}{2}} = \l(x_{+}'\,, \q')\,. 
\end{align}
This completes the proof.

Now that we have determined the transformation properties of the two-planes $\mathfrak{T}$ under both finite superconformal and finite AdS transformations, let us introduce the matrix two-point function 
\begin{align}
\cE(\mathfrak{T}_{1}\,,\mathfrak{T}_2) = \mathfrak{T}_{1}^{\dag} \O \mathfrak{T}_{2}\,,
\end{align}
for two two-planes $\mathfrak{T}_1$ and $\mathfrak{T}_2$. 
Given the null condition \eqref{null con}, it follows that 
\begin{align}
\cE(\mathfrak{T}_{1}\,,\mathfrak{T}_{1}) = 0\,. 
\end{align}
The two-point function $\cE(\mathfrak{T}_{1}\,,\mathfrak{T}_{2})$ transforms homogeneously under superconformal transformations
\begin{align}
\cE(\mathfrak{T}_1'\,,\mathfrak{T}_2') = R^{\dag}(\hat{g}\,,\mathfrak{T}_1)\cE(\mathfrak{T}_1\,,\mathfrak{T}_2)R(\hat{g}\,,\mathfrak{T}_2)\,,
\end{align}
and under AdS transformations
\begin{align} \label{ads 2p trf} 
\cE(\mathfrak{T}_1'\,,\mathfrak{T}_2') = N^{\dag}(g\,,\mathfrak{T}_1)\cE(\mathfrak{T}_1\,,\mathfrak{T}_2)N(g\,,\mathfrak{T}_2)\,. 
\end{align}

Associated with $\cE(\mathfrak{T}_1\,,\mathfrak{T}_2)$ is the two-point function 
\begin{align}
\o(\mathfrak{T}_1\,,\mathfrak{T}_2) = \det \cE(\mathfrak{T}_1\,,\mathfrak{T}_2)\,,
\end{align}
with the superconformal transformation law 
\begin{align}
\o(\mathfrak{T}'_1\,,\mathfrak{T}'_2) = \det R^{\dag}(\hat{g}\,,\mathfrak{T}_1) \det R(\hat{g}\,,\mathfrak{T}_2) \, \o(\mathfrak{T}_1\,,\mathfrak{T}_2)\,.
\end{align}
Given the relation \eqref{ads 2p trf} and the fact that $N(g\,,\mathfrak{T}) \in \sSL(2\,,\mathbb{C})$, it follows that $\o(\mathfrak{T}_1\,,\mathfrak{T}_2)$ is invariant under the AdS transformations. 

If we restrict our attention to the AdS supergroup only, we can introduce chiral and antichiral two-point functions 
\bsubeq
\begin{align}
\cE_{+}(\mathfrak{T}_1\,,\mathfrak{T}_2) &= \mathfrak{T}_1^\sT \frak{J} \mathfrak{T}_2\,, 
\\
\cE_{-}(\mathfrak{T}_1\,,\mathfrak{T}_2) &= (\star \mathfrak{T}_1)^{\sT}\frak{J}\star \mathfrak{T}_2\,,
\end{align}
\esubeq
and 
\bsubeq
\begin{align}
\o_{+}(\mathfrak{T}_1\,,\mathfrak{T}_2) &= \det \cE_{+}(\mathfrak{T}_1 \,, \mathfrak{T}_2)\,,
\\
\o_{-}(\mathfrak{T}_1\,,\mathfrak{T}_2) &= \det \cE_{-}(\mathfrak{T}_1\,,\mathfrak{T}_2)\,.
\end{align}
\esubeq
The two-point functions $\o\,,\o_{+}\,,$ and $\o_{-}$ are invariant under the AdS transformations \eqref{finite ads trf}.
However, under equivalence transformations \eqref{gl equiv}, they scale as 
\bsubeq \label{chiral anti chiral scaling}
\begin{align}
\o(\mathfrak{T}_1\,,\mathfrak{T}_2) &\rightarrow \det R_{1}^{\dag} \det R_2 ~ \o(\mathfrak{T}_1\,,\mathfrak{T}_2)\,, 
\\
\o_{+}(\mathfrak{T}_1\,,\mathfrak{T}_2) &\rightarrow \det R_1 \det R_2 ~ \o_{+}(\mathfrak{T}_1\,,\mathfrak{T}_2)\,, 
\\
\o_{-}(\mathfrak{T}_1\,,\mathfrak{T}_2) &\rightarrow \det R_{1}^{\dag} \det R_{2}^{\dag}~\o_{-}(\mathfrak{T}_1\,,\mathfrak{T}_2)\,. 
\end{align}
\esubeq

Making use of the above analysis, we can construct a two-point function that is invariant under both the AdS transformations \eqref{finite ads trf} and arbitrary equivalence transformations of the form \eqref{gl equiv}, as follows
\begin{align}
\tilde{\o}(\mathfrak{T}_1\,,\mathfrak{T}_2) = \frac{\o(\mathfrak{T}_1\,,\mathfrak{T}_2)}{\sqrt{\o_{-}(\mathfrak{T}_1\,,\mathfrak{T}_1)\,
\o_{+}(\mathfrak{T}_2\,,\mathfrak{T}_2)}}\,.
\end{align}
Choosing $\mathfrak{T}_1 = \mathfrak{T}$ and  $\mathfrak{T}_2 = \mathfrak{T} + \rd \mathfrak{T}$ allows us to obtain the AdS supersymmetric interval defined by 
\begin{align}
\rd s^{2} = \tilde{\o}(\mathfrak{T}\,, \mathfrak{T} + \rd \mathfrak{T})\,.
\end{align}

Let us evaluate $\tilde{\o}(\mathfrak{T}\,, \mathfrak{T} + \rd \mathfrak{T})$ in the north chart. 
We find 
\begin{align}
\cE(\mathfrak{T}\,, \mathfrak{T} + \rd \mathfrak{T}) = - \ri |\l|^{2} \P^{a} \tilde{\s}_{a}\,, \qquad \P^{a} = \rd x^{a} + \ri(\q\s^{a}\rd\bar{\q} - \rd\q \s^{a} \bar{\q})\,, 
\end{align}
where $\P^{a}$ is the Volkov-Akulov one-form \cite{VA,AV}.
We end up with the supermetric 
\begin{align}
\rd s^{2} = \l^{2} \bar{\l}^{2} \P^{a} \P^{b}\eta_{ab}\,.
\end{align}
which is AdS-invariant.

\section{The supertwistor realisation of $ \text{AdS}^{4|8} \times {\mathbb F}_1(2) $} \label{section6}

In this section we will develop a supertwistor realisation for the flag superspace \eqref{4D-AdS-flag}. 
Such a realisation necessarily makes use of odd supertwistors, for which our conventions are described in section \ref{section4}. The supertwistor realisation for the flag superspace  $\overline{\mathbb M}^{4|8}  \times {\mathbb F}_1(2) $ was given in \cite{Kuzenko:2006mv}, and here we will build on that construction.

Our starting point is the space of quadruples $\{T^\m, \X^+, \X^- \}$
consisting of two even supertwistors $T^\m$ and 
two odd supertwistors $\X^\pm$ such that (i) the bodies of 
$T^\m$ are linearly independent four-vectors; 
and (ii) the bodies of $\X^\pm$ are linearly independent two-vectors.
These supertwistors
are further required to obey the relations 
\bsubeq \label{harmonic con}
\begin{align}
\braket{T^{\m}}{T^{\n}}_{\O} 
&= 0\,, \qquad \braket{T^{\m}}{\X^{\pm}}_{\O} = 0\,, 
\\
\braket{T^{\m}}{T^{\n}}_{\frak{J}} &= \ell \ve^{\m\n}\,, 
\end{align}
\esubeq
and are defined modulo the equivalence  relation
\bea \label{harmonic equiv}
(\X^-,\X^+, T^\m)\sim   (\X^-,\X^+, T^\n) \,
\left(
\begin{array}{cc||c}
a~& 0~& 0 \\
b~& c~  &0  \\  \hline \hline
 \r^-_\n~ & \r^+_\n ~&R_\n{}^\m 
\end{array}
\right) ~,\quad 
\left(
\begin{array}{cc||c}
a~& 0~& 0 \\
b~& c~  &0  \\  \hline \hline
 \r^- ~ & \r^+  ~&R 
\end{array}
\right)
 \in \sGL(2|2;\mathbb{C})~,
\eea
with $\r^\pm_\n$  anticommuting complex parameters.
No symplectic condition is imposed on the odd supertwistors $\X^{\pm}\,.$
As above, one can work with normalised two-planes by fixing a particular value of $\ell$. Then, the gauge freedom \eqref{harmonic equiv} is reduced such that the $2 \times 2$ matrix $R \in \sSL(2\,,\mathbb{C})\,.$ In what follows we will set $\ell = 1\,.$ 

The even and odd supertwistors can be represented as 
\begin{align}
\mathfrak{T} = \left(\begin{array}{c}
F \\ 
G 
\\
\hline \hline 
\vf
\end{array}\right)\,, 
\qquad
\X^{\pm} = \left(\begin{array}{c}
\x^{\pm} \\ 
\psi^{\pm} \\ 
\hline \hline 
V^{\pm}
\end{array}\right)\,.
\end{align}
Then, the conditions \eqref{harmonic con} imply the following 
\begin{align}
F^{\dag}\psi + G^{\dag}\x - \vf^{\dag}V = 0\,,
\end{align}
in addition to those we encountered in the previous section, eq. \eqref{block con}. 

In the north chart, where $\det F \neq 0\,,$ the supertwistors can be chosen to take the form 
\begin{align} \label{nc harmonic 2p}
\mathfrak{T}  = \l \left(\begin{array}{c}
\id_{2} \\ 
-\ri \tilde{x}_{+} \\ \hline \hline 
2\q
\end{array}\right)\,, 
\qquad 
\X^{\pm} = \left(\begin{array}{c}
0 \\ 
2 \bar{\q}^{\pm} 
\\ \hline \hline
v^{\pm}
\end{array}\right)\,,
\end{align}
with 
\bsubeq
\begin{align}
v^{\pm} = (v^{\pm}_{i})\,, \qquad \bar{\q}^{\pm} = \left(\bar{\q}^{\pm \ad}\right)\,,
\end{align}
and
\bea
\det  \Big(v_i{}^- , v_i{}^+ \Big) =
 v^{+i} \,v^-_i \neq 0~,
\qquad v^{+i} = \ve^{ij} \,v^+_j~.
\non
\eea
\esubeq
The orthogonality conditions $\braket{T^{\m}}{\X^{\pm}}_{\O} = 0$ imply 
\begin{align}
\bar{\q}^{\pm\ad} = \bar{\q}^{\ad i}v_{i}^{\pm}\,.  
\end{align}
The complex harmonic variables $v^\pm_i$ 
in \eqref{nc harmonic 2p} are still defined  modulo arbitrary
transformations of the form 
\bea
\Big(v_i{}^- , v_i{}^+ \Big)  ~\to ~
\Big(v_i{}^- , v_i{}^+ \Big) 
\,\tilde{\bm r}~,
\qquad 
\tilde{\bm r}= \left(
\begin{array}{cc}
 a  & 0\\
b &      c 
\end{array}
\right) \in \sGL(2,{\mathbb C})~.
\label{equivalence22}
\eea
We see that the complex harmonic variables $v^{\pm}$ parametrise  ${\mathbb F}_1(2)$ as described in Section \ref{section2}. 
It follows that the set $\{T^{\m}\,, \X^{-}\,, \X^{+}\}$ constitutes a supertwistor realisation of the AdS flag superspace \eqref{4D-AdS-flag}. 

Let us make use of the equivalence relation \eqref{equivalence22} to impose the condition
\begin{subequations}
\bea 
v^{+i} \,v^-_i =1~.
\label{unimod}
\eea
The harmonics then obey the identity 
\begin{align}\label{harm completeness}
v^{+}_{i}v^{-}_{j} - v^{+}_{j}v^{-}_{i} = \ve_{ij}\,.
\end{align}
\end{subequations}
As explained in Section \ref{section2}, the gauge freedom
\eqref{equivalence22} allows one to represent 
any infinitesimal transformation of the harmonics in the form \eqref{delta v}.

To determine the action of the AdS supergroup on the harmonic variable $v^{+}$, it is useful to begin with an infinitesimal superconformal transformation \eqref{scf alg}.
It can be shown that
\bsubeq
\begin{align}
\d v_{i}^{+} &= -\tilde{\L}^{++} v_{i}^{-}\,,
\end{align}
where $\tilde{\L}^{++}$ is expressed as
\be
\tilde{\L}^{++} =\L^{ij} \,v^+_i v^+_j - 4 \, {\rm i}\,\q^+ \,b \,{\bar \q}^+
- 4( \q^+ \eta^+ -{\bar \q}^+ {\bar \eta}^+ ) ~,
\ee 
\esubeq
see \cite{Kuzenko:2006mv} for the derivation.
Making use of the derivatives $D_{\a}^{+}$ and $\bar{D}_{\ad}^{+}$ defined by eq. \eqref{1.17}, we can see that
\begin{align}
D_{\a}^{+}\tilde{\L}^{++} = \bar{D}_{\ad}^{+}\tilde{\L}^{++} = 0\,.
\end{align}
The variations of $\q^{+\a}$ and $\bar{\q}^{+\ad}$ are given by 
\begin{align}
\d \q^{+\a} = \d \q^{\a i} v_{i}^{+} - \tilde{\L}^{++}\q^{\a i} v_{i}^{-}\,, \qquad \d \bar{\q}^{+\ad} = \d \bar{\q}^{\ad i} v_{i}^{+} - \tilde{\L}^{++}\bar{\q}^{\ad i}v_{i}^{-}\,,
\end{align}
where $\d \q^{\a i}$ is given by \eqref{nc chiraltra}. 
Further, they satisfy the property 
\begin{align} \label{analytic variation}
D^{+}_{\b}\d\q^{+}_{\a} = \bar{D}^{+}_{\bd}\d\q^{+}_{\a} = 0\,. 
\end{align}
Now, one can single out the AdS transformations by imposing the constraints \eqref{ads translation con} and \eqref{ads so(n) con}. 

Finally, we comment on the analytic bosonic coordinates 
\begin{align}
y^{a} = x^{a} - 2\ri\q^{(i}\s^{a}\bar{\q}^{j)}v_{i}^{+}v_{j}^{-}\,, \qquad D_{\a}^{+}y^{a} = \bar{D}_{\ad}^{+}y^{a} = 0\,. 
\end{align}
It can be shown that, under an infinitesimal transformation of the AdS supergroup, the variation $\d y^{a}$ satisfies 
\bsubeq \label{analytic y variations}
\begin{align}
D_{\a}^{+} \d y^{a} &= D_{\a}^{+} \d \left( x_{-}^{a} - \ri \q^{+}\s^{a}\bar{\q}^{-}\right) = 0 \,,
\\
\bar{D}_{\ad}^{+} \d y^{a} &= \bar{D}_{\ad}^{+} \d \left( x_{+}^{a} - \ri\q^{-}\s^{a}\bar{\q}^{+} \right) = 0\,.
\end{align}
\esubeq
In order to prove the relations \eqref{analytic y variations}, we make use of the identities \eqref{harm completeness} and \eqref{analytic variation}, as well as the fact that $\d x_{+}^{a}$, defined by eq. \eqref{nc chiraltra} subject to the conditions \eqref{ads translation con}, is chiral, $\bar{D}_{\ad}^{i} \d x_{+}^{a} = 0\,.$
It follows that the analytic subspace parametrised by the variables 
\begin{align}
\z = (y^{a}\,, \q^{+\a}\,, \bar{\q}^{+}_{\ad}\,, v^{+}_{i}\,, v^{-}_{i})\,, \qquad D_{\a}^{+}\z = \bar{D}_{\ad}^{+}\z = 0\,,
\end{align}
is invariant under the AdS supergroup. 


\section{Conclusion} 

When dealing with off-shell $\cN=1$ and $\cN=2$ supersymmetric field theories in AdS$_4$
(see 
\cite{IS, KT-M08, BK11, Festuccia:2011ws, Butter:2011zt, Butter:2011kf, Butter:2012jj, Aharony:2015hix} for an incomplete list of references) 
one usually makes use of  {\it local} superspace differential geometry.
In the $\cN$-extended case, the algebra of covariant derivatives for AdS$^{4|4\N}$ is given by the graded commutation relations \cite{Koning:2023ruq, Koning:2024iiq}
\begin{subequations} 
	\label{2.170}
	\bea
	\{ \cD_\a^i , \cD_\b^j \}
	&=&
	4 S^{ij}  M_{\a\b} 	- 4 \ve_{\a\b}S^{k[i}  \mathbb{J}^{j]}{}_{k}
	~,
	\\
	\{ \bar{\cD}^\ad_i , \bar{\cD}^\bd_j \}
	&=&
	- 4 \bar{S}_{ij}  \bar{M}_{\ad\bd} 	+ 4 \ve^{\ad \bd} \bar{S}_{k[i}  \mathbb{J}^{k}{}_{j]}
	~,
	\\
	\{ \cD_\a^i , \bar{\cD}^\bd_j \}
	&=&
	- 2 \ri \d_j^i\cD_\a{}^\bd
	~,
	\\
	{[} \cD_\a^i, \cD_{\bb}{]}&=& 
	- \ri \ve_{\a \b} S^{ij} \bar{\cD}_{\bd j}
	~,
	\qquad
	{[} \bar{\cD}^\ad_i, \cD_{\bb}{]}= 
	\ri \d_\bd^\ad \bar{S}_{ij} \cD_\b^j
	~, 
	\\
	\big [ \cD_\aa , \cD_\bb \big] &=& - 2|S|^2
	(\ve_{\a \b} \bar{M}_{\ad \bd} + \ve_{\ad \bd} M_{\a \b})~,\qquad |S|^2: = \frac{1}{\cN}S^{ij} \bar{S}_{ij}>0~.
	\eea
\end{subequations}
Here ${\mathbb J}^i{}_j$ denotes the $\sSU(\cN)$ generator, $S^{ij} = S^{ji}$ is a covariantly constant curvature  tensor, $\bar S_{ij} := \overline{S^{ij}}$,  
 When $\cN>1$, the constraint $\cD_A S^{jk} = 0$ implies the following integrability condition
\begin{align}
	\d_{(k}^{[i} S^{j]m} \bar{S}_{l)m} = 0 \quad \implies \quad 
	S^{ik} \bar{S}_{jk} = \frac{1}{\cN} \d^i_j S^{kl} \bar{S}_{kl}
	~.
\end{align}

As demonstrated in \cite{Koning:2024iiq}, performing a local $\sU(\cN)_R$ transformation allows one to bring $S^{ij}$ to the form 
\begin{align}
	S^{ij} = \d^{ij} S ~. 
\end{align}
Now, the condition $\cD_A S^{jk} = 0$ tells us the $\sSU(\cN)_R$ connection involves only the operators 
\begin{align}
	\mathcal{J}^{ij} := - 2 \d^{k[i} \mathbb{J}^{j]}{}_k = - \cJ^{ji}\,,
\end{align}
which generate the group $\sSO(\cN)$. The algebra of covariant derivatives then takes the form:
\begin{subequations} 
	\label{AdS}
	\bea
	\{ \cD_\a^i , \cD_\b^j \}
	&=&
	4 S \d^{ij}  M_{\a\b} + 2 \ve_{\a\b} S \cJ^{ij}
	~,
	\\
	\{ \bar{\cD}^\ad_i , \bar{\cD}^\bd_j \}
	&=&
	- 4 \bar{S} \d_{ij}  \bar{M}^{\ad\bd} - 2 \ve^{\a\b} \bar{S}  \cJ_{ij}
	~,
	\\
	\{ \cD_\a^i , \bar{\cD}^\bd_j \}
	&=&
	- 2 \ri \d_j^i\cD_\a{}^\bd
	~,
	\\
	{[} \cD_\a^i, \cD_{\bb}{]}&=& 
	- \ri \ve_{\a \b} S \bar{\cD}_{\bd}^i
	~,
	\qquad
	{[} \bar{\cD}^\ad_i, \cD_{\bb}{]}= 
	\ri \d_\bd^\ad \bar{S} \cD_{\b j}
		~, 
		\\
		\big [ \cD_\aa , \cD_\bb \big] &=& - 2 |S|^{2} (\ve_{\a \b} \bar{M}_{\ad \bd} + \ve_{\ad \bd} M_{\a \b})~.
		\eea
\end{subequations}

In a previous series of papers \cite{Kuzenko:2021vmh,Koning:2023ruq,Koning:2024iiq}, we have developed the global approach to  the AdS$^{4|4\N}$ supergeometry. The novelty of the present paper is that we have reformulated the global realisation of AdS$^{4|4\N}$ by explicitly embedding the AdS supergroup in the superconformal group. The virtue of this approach is that: (i) it shows how the AdS superspace arises as a certain open domain of $\overline{\mathbb{M}}{}^{4|4\N}$; (ii) it allows us to read off the superconformal and isometry transformation rules for AdS$^{4|4\N}$ in terms of those known for $\overline{\mathbb{M}}{}^{4|4\N}$; and (iii) it proves most suitable for developing a global realisation of the flag superspace \eqref{4D-AdS-flag}. 

In a recent work \cite{Ivanov:2025jdp} Ivanov and Zaigraev derived the AdS isometry transformations on the analytic subspace of the $\cN=2$ AdS harmonic superspace. Their construction was based on the following two inputs: (i)  the known $\cN=2$ superconformal transformations in the analytic subspace of harmonic superspace \cite{GIOS-conf}; and (ii)  the known embedding of the $\cN$-extended AdS superalgebra $\mathfrak{osp}(\cN|4;{\mathbb R})$ in  the $\cN$-extended superconformal algebra  $\mathfrak{su}(2,2|\cN)$ \cite{GGRS,Bandos:2002nn}. 
Their analysis was limited to an open domain of the $\cN=2$ AdS harmonic superspace, and no discussion of its global structure was given. 
In our approach, the AdS isometry transformations acting on the analytic subspace of the $\cN=2$ AdS harmonic superspace are readily derived from the supertwistor formulation given.

Let us now consider the algebra \eqref{AdS} in the $\N=2$ case. The presence of the $\sSO(2)$ generator suggests that it is quite natural to consider an AdS superspace of the form AdS$^{4|8} \times S^{1}$, although this superspace does not allow the action of the superconformal group.
Making use of our construction in the present paper, one can readily derive a supertwistor realisation of this superspace.\footnote{See \cite{Koning:2024vdl} for a similar story in five dimensions.} 
To do so, we introduce the space of triples $\{T^{\m}\,, \X\}$ consisting of two even supertwistors $T^{\m}$ and a single real odd supertwistor $\X = \star{\X}$ such that: (i) the bodies of $T^{\m}$ are linearly independent four-vectors; and (ii) the body of $\X$ is non-vanishing. These supertwistors are required to obey the relations 
\bsubeq
\begin{align}
\braket{T^{\m}}{T^{\n}}_{\O} &= 0\,, \qquad \braket{T^{\m}}{\X}_{\O} = 0\,,
\\
\braket{T^{\m}}{T^{\n}}_{\frak{J}} &= \ell \ve^{\m\n}\,,
\end{align}
\esubeq
and are defined modulo the equivalence relation 
\begin{align}
(\X\,, T^{\m}) \sim (\X\,, T^{\n})\left(\begin{array}{c||c}
a & ~0~ \\
\hline \hline 
~0~ & R_{\n}{}^{\m}
\end{array}\right)\,, \qquad a \in \mathbb{R} - \{0\}\,, \quad  R \in \sGL(2\,,\mathbb{C})\,. 
\end{align}
The superspace obtained is seen to be 
\begin{align}
\text{AdS}^{4|8} \times \mathbb{R}P^{1} \simeq \text{AdS}^{4|8} \times S^{1}\,. 
\end{align}
\\

\noindent
{\bf Acknowledgements:}\\
We are grateful to Emmanouil Raptakis for comments on the manuscript. 
 This work  is supported in part by the Australian Research Council, project DP230101629.

\appendix 

\section{$\cN=2$ conformal Killing supervector fields}\label{appendixA}

This appendix is devoted to a brief review of the $\cN=2$ conformal Killing supervector fields. Our presentation is inspired by \cite{Kuzenko:1999pi} and follows \cite{KRT-M_N=2}.

An infinitesimal superconformal  transformation 
\bea
 z^A \to    z^A  + \d z^A~, \qquad
\d z^A =\x  \,z^A = \Big(\x^a +\ri (\x_i\s^a \bar \q^i - \q_i \s^a \bar \x^i), \x^\a_i, \bar \x_\ad^i \Big)
\eea
is generated by 
a {\it conformal Killing supervector field}
\bea
\x = \x^b  \pa_b + \x^\b_j D_\b^j
+ {\bar \x}_{\bd}^j  {\bar D}^{\bd}_j = {\overline \x}
~.
\eea 
The defining property of $\x$ is 
\begin{align}
	[\x , D_\a ^i] = - (D_\a^{i} \x^\b_j) D_\b^j~.
\end{align}
This condition implies the relations 
\bea
{\bar D}^{ \ad }_i \x^\b_j =0~, \qquad
{\bar D}^{ \ad }_i \x^{ \bd \b} = 4{\rm i} \, \ve^{\ad \bd} \x^\b_i \quad \implies \quad
\x^\a_i = -\frac{\ri }{8} \bar D_{\ad i} \x^{\ad \a }  
\eea
and their complex conjugates, 
and therefore
\bea
{\bar D}_{(\a i} \x_{\b) \bd } =0~, \qquad {\bar D}_{(\ad }^i \x_{\b \bd )} =0 
\quad \implies \quad \pa_{(\a (\ad} \x_{\b) \bd)}=0~.
\eea
It then follows that
\bea
[\x , D_\a^i ]
= - K_\a{}^\b [\x] D_\b^i - \hf \bar{\s}[\xi] D_\a^i - \L^{i}{}_{j}[\xi] D_{\a}^j~.
\label{4Dmaster2N=2} 
\eea
Here we have introduced the chiral Lorentz
$K_{\b\g}[\x]$ and 
super-Weyl $\s[\x]$ parameters, 
as well as the $\sSU(2)_R$ parameter $K^{ij}[\xi]$ defined by 
\begin{subequations} 
\label{2.7}
\bea
K_{\a\b}[\x]&=& \hf D_{(\a}^i \x_{\b)i}=K_{\b \a}[\xi]~, \qquad \bar D^\ad_i K_{\a\b}[\x] =0~, \\
\s [\x] &= & \frac{1}{2} \bar{D}^\ad_i \bar{\xi}_\ad^i~, \qquad \qquad \qquad \quad \;\;\;\;\bar{D}^\ad_i \s[\xi] = 0~, \label{2.7b} \\
\L^{ij}[\xi] &=& - \frac{\ri}{16} [D^{(i}_\a , \bar{D}_\ad^{j)}] \xi^{\aa} = \L^{ji}[\xi]~,
\quad 
\overline{\L^{ij}[\xi] } = \L_{ij}[\xi] ~. \label{2.7c}
\eea
\end{subequations}
We recall that the Lorentz parameters with vector and spinor indices are related to each other as follows:
$K^{bc}[\x] = (\s^{bc})_{\b\g}K^{\b\g}[\x] - (\tilde{\s}^{bc})_{\bd\gd}\bar K^{\bd\gd}[\x] $.
The parameters in \eqref{2.7} obey several first-order differential properties:
\begin{subequations}
\bea
D^i_\a {\L}^{jk} [\x]&=&
\ve^{i(j} D_\a^{k)} 
\s [\x]~,
\label{an1N=2} \\
D^i_\a K_{\b\g} [\x] &=& -\ve_{\a(\b} D^i_{\g)} \s[\x]~,
\eea
\end{subequations}
and therefore 
\begin{subequations}
\bea
D^{(i}_\a \L^{jk)} [\x]&=& {\bar D}^{(i}_{\ad} \L^{jk)} [\x]=0~, 
\label{L-an} \\
D^i_\a D^j_\b \s[\x] &=& 0~.
\eea
\end{subequations}

The superconformal transformation law of a primary tensor superfield (with suppressed indices) is
\begin{subequations}\label{FlatPrimaryMultiplet}
\bea
\d_\x U &=& \cK[\x] U,  \\
 \cK[\x] &=& \x +\hf K^{ab}[\x]M_{ab} + \L^{ij}[\x] J_{ij} 
+p\s[\x] +q\bar \s[\x] ~.
\eea
\end{subequations}
Here the generators $M_{ab}$ and $J_{ij}$ act on the Lorentz and $\sSU(2) $ indices of $U$, respectively. The parameters $p$ and $q$ are related to the dimension (or Weyl weight) $w$ and $\sU(1)_R $ charge $c$ of $U$ as $ p+q =w$ and $p-q = - \hf c$. 

The most general $\cN=2$ conformal Killing supervector field has the form
\begin{subequations} \label{chiraltra}
	\bea
	\x_+^{\ad\a} &=& a^{\ad\a}  +\hf (\D +{\bar \D})\, x_{+}^{\ad\a}
	+{\bar K}^\ad{}_\bd  \,x_{+}^{\bd \a} +x_{+}^{\ad\b}K_\b{}^\a 
	-x_{+}^{\ad \b} b_{\b \bd} x_{+}^{\bd \a} \non \\
	&& \qquad +4{\rm i}\, {\bar \e}^{\ad i}  \q^{\a}_i - 4 x_{+}^{\ad \b} \eta_\b^i \q^\a_i ~,
	\\
	\x^\a_i &=& \e^\a_i + \hf \bar{\D} \q^\a_i + \q^\b_i K_\b{}^\a + \L_{i}{}^{j} \q^{\a}_j
	-  \q^\b_i b_{\b\bd}   x_{+}^{\bd \a} \non \\	
	&& \qquad -{\rm i}\,{\bar \eta}_{\bd i} x_{+}^{\bd \a} - 4 \q^\b_i \eta_\b^j \q^\a_j~,~~~~
	\eea
\end{subequations}
where we have introduced the complex four-vector
\be
\x_+^a = \x^a + 2\ri \x_i \s^a  {\bar \q}^i~, 
\qquad \bar \x^a =\x^a~,
\ee
along with the complex bosonic coordinates $x_{+}^a = x^a +\ri \q_i \s^a \bar \q^i$ 
of the chiral subspace of ${\mathbb M}^{4|8}$. 
The constant bosonic parameters in \eqref{chiraltra}
correspond to the spacetime translation ($a^{\ad \a}$), 
Lorentz transformation ($K_\b{}^\a,~{\bar K}^{\ad}{}_{\bd})$,
$\sSU(2)_R$ transformation ($\L^{ij} = \L^{ji}$),
special conformal transformation
($ b_{\a \bd}$), and  combined scale and $\sU(1)_R$ transformations 
($\D =\t - 2 \ri \vf$). The constant fermionic parameters in \eqref{chiraltra}
correspond to the $Q$-supersymmetry ($\e^\a_i$) and $S$-supersymmetry 
($\eta_i^\a$) transformations. The constant parameters $K_{\a\b}$, $\L^{ij}$ and $\D$ are obtained 
from $K_{\a\b}[\x]$, $\L^{ij}[\x]$ and $\s[\x]$, respectively,  by setting $z^A=0$.

In the case of the $Q$-supersymmetry transformation,  when the only non-vanishing parameters in 
\eqref{chiraltra} are $\e^\a_i$ and its conjugate, it holds that the descendants $K^{ab}[\x]$, $ \L^{ij}[\x] $ and $\s[\x] $ vanish, 
and the transformation law \eqref{FlatPrimaryMultiplet} takes the universal form 
\bea
\d_\e U = \Big( 2\ri \big( \q_i \s^a\bar \e^i - \e_i \s^a \bar \q^i\big)  \pa_b + \e^\a_i D_\a^i
+ {\bar \e}_{\ad}^i  {\bar D}^{\ad}_i \Big) U
=:  \big( \e^\a_i Q^i_\a + \bar \e_\ad^i \bar Q^\ad_i \big) U~.
\eea


\section{Another similarity transformation for the AdS supergroup}\label{appendixB}

In section \ref{section new ads group}, we described an isomorphic realisation of the AdS supergroup, denoted $\sOSp_0(\N|4;\mathbb{R})_{\frak{U}}$, which was useful for our applications in this paper. 
It turns out that there is another unitary supermatrix which relates the two realisations of the AdS supergroup $\sOSp_{0}(\N|4;\mathbb{R})$ and $\sOSp_0(\N|4;\mathbb{R})_{\frak{U}}\,.$
Below, we will describe this supermatrix and how it is related to that of eq. \eqref{u def}.

Let us introduce the supermatrix $\frak{N}$ defined as 
\begin{align}
\frak{N} = \left(\begin{array}{c||c}
\frak{n} & 0 \\
\hline \hline 
0 & \id_{\N}
\end{array}\right)\,, \qquad 
\frak{n }= \frac{\re^{-\ri\p/4}}{\sqrt{2}} \left(\begin{array}{c|c}
\id_{2} & - \ve \\ 
\hline
\ri\ve & - \ri\id_{2} 
\end{array}\right)\,.
\end{align}
The supermatrix $\frak{N}$ enjoys the properties
\bsubeq
\begin{align}
\frak{N}^{\dag} &= \frak{N}^{-1}\,, 
\\
\frak{N}^{\dag}\mathbb{J}\frak{N} &= -\ri\O\,,
\\
\frak{N}^{\sT}\mathbb{J}\frak{N} &= \frak{J}\,. 
\end{align}
\esubeq
It turns out that, for every $f \in \sOSp_{0}(\N|4;\mathbb{R})$, the supermatrix defined by 
\begin{align}
g = \frak{N}^{-1}f\frak{N}
\end{align}
belongs to $\sOSp_0(\N|4;\mathbb{R})_{\frak{U}}\,.$

As the supermatrices $\frak{N}$ and $\frak{U}$ both take us to the realisation $\sOSp_0(\N|4;\mathbb{R})_{\frak{U}}$, they must be related to each other in the following way 
\bsubeq \label{u to n}
\begin{align} 
\frak{U} = \frak{N} \frak{S}\,,
\end{align}
where $\frak{S}$ satisfies the following properties 
\begin{align}
\frak{S}^{\dag} = \frak{S}^{-1}\,, \qquad \frak{S}^{\dag}\O \frak{S} = \O\,, \qquad \frak{S}^{\sT}\frak{J}\frak{S} = \frak{J}\,.
\end{align}
\esubeq
It can be shown that the solution to eq. \eqref{u to n} takes the form 
\begin{align}
\frak{S} = \left(\begin{array}{c|c}
\frak{s} & ~0~ \\
\hline \hline 
~0~ & \id_{\N}
\end{array}\right)\,, \qquad 
\frak{s} = \frac{1}{\sqrt{2}}\left(\begin{array}{c|c}
\ve & \ri \ve \\ 
\hline 
\ri \ve & \ve
\end{array}\right)\,.
\end{align}


\section{The Killing supervectors of AdS$^{4|4\N}$} \label{appendixC}

In this appendix we will provide an alternative derivation of the constraints \eqref{ads translation con} and \eqref{ads so(n) con}, making use of the $\sSU(\N)$ superspace formulation for AdS$^{4|4\N}$ developed in \cite{Koning:2023ruq, Koning:2024iiq}.

AdS$^{4|4\N}$ is parametrised by local coordinates $z^{M} = (x^{m}\,, \q_{\imath}^{\m}\,, \bar{\q}^{\imath}_{\dmu})$. Its covariant derivatives $\cD_{A} = (\cD_{a}\,, \cD_{\a}^{i}\,, \bar{\cD}_{i}^{\ad})$ take the form 
\begin{align}
\cD_{A} = E_{A} + \frac{1}{2}\O_{A}{}^{cd}M_{cd} + \F_{A}{}^{i}{}_{j}\mathbb{J}^{j}{}_{i}\,.
\end{align}
Here, $M_{cd} = - M_{dc}$ are the Lorentz generators, and $\mathbb{J}^{i}{}_{j}$ are the $\sSU(\N)$ generators. 
In a conformally flat frame, the covariant derivatives are given by the following expressions 
\begin{subequations}
	\label{AdSBoost}
	\begin{align}
	\cD_\a^{i}&= \re^{\frac{\cN-2}{2\cN} \s + \frac 1 \cN \bar{\s}} \Big( D_\a^i+ D^{\b i}\s M_{\a \b} + D_{\a}^j\s \mathbb{J}^{i}{}_j \Big) ~, 
	\\ 
	\bar{\cD}_{i}^{\ad}&=\re^{\frac{1}{\cN} \s + \frac{\cN-2}{2\cN} \bar{\s}} \Big( \bar{D}^\ad_i-\bar{D}_{ \bd i} \bar{\s} \bar{M}^{\ad \bd} - \bar{D}^{\ad}_j \bar{\s} \mathbb{J}^{j}{}_i \Big)~,
	\\
	\cD_\aa &= \re^{\hf \s + \hf \bar{\s}} \Big(\partial_\aa + \frac{\rm i}{2} D^i_{\a} \s \bar{D}_{\ad i} + \frac{\rm i}{2} \bar{D}_{\ad i} \bar{\s} D_{\a}^i + \hf \Big( \partial^\b{}_\ad (\s + \bar \s ) - \frac{\ri}{2} D^{\b i} \s \bar{D}_{\ad i} \bar{\s} \Big) M_{\a \b} \non \\ & + \hf \Big( \partial_{\a}{}^\bd (\s + \bar{\s}) + \frac{\ri}{2} D_{\a}^i \s \bar{D}^{\bd}_i \bar{\s} \Big) { \bar M}_{\ad \bd} 
	- \frac \ri 2 D_\a^i \s \bar{D}_{\ad j} \bar{\s} \mathbb{J}^j{}_i \Big) ~, 
	\end{align}
\end{subequations}
where $D_{A} = (\partial_{a}\,, D_{\a}^{i}\,, \bar{D}_{i}^{\ad})$ are the flat $\N$-extended covariant derivatives, and $\s$ is a chiral superfield, $\bar{D}^{\ad}_{i}\s = 0$, satisfying the constraints 
\bsubeq
\begin{align} \label{ads constraints 1}
\quad D_{(\a}^{[i} D_{\b)}^{j]} \re^\s = 0~,
		\\
[D_\a^i,\bar{D}_{\ad i}] \re^{\frac{\cN}{2}(\s + \bar{\s})} = 0 \label{ads constraints 2}
		~.
\end{align}
\esubeq
In the $\N=1$ case, the constraint \eqref{ads constraints 1} should be replaced with 
\begin{align} \label{ads n=1 constraints}
-\frac{1}{4}\re^{2\bar{\s}}D^{2}\re^{-\s} = \text{const}\,. 
\end{align}
The constraints \eqref{ads constraints 1}, \eqref{ads constraints 2}, and \eqref{ads n=1 constraints} have been solved in \cite{Buchbinder:1998qv,Butter:2012jj,KT-M08} in the $\N=1$ and $\N=2$ cases, for both stereographic coordinates and Poincar\'e coordinates. 
In the $\N$-extended case, they are solved in \cite{Koning:2024iiq} for both realisations. 
For the stereographic solution, $W = \re^{-\s}$ takes the form 
\begin{align}\label{ads compensator}
W &= (1 - \frac{s^{ij}\bar{s}_{ij}}{4\N}x_{+}^{2} + s^{ij}\q_{ij})^{-1}\,,
\end{align}
where $s^{ij}$ satisfies the properties 
\begin{align}
s^{ij} = s^{ji}\,, \qquad s^{ik}\bar{s}_{kj} = |s|^{2}\d^{i}_{j}\,, \qquad \bar{s}_{ij} := \overline{s^{ij}}\,. 
\end{align}
The superfield $W$ plays the role of the compensator for AdS$^{4|4\N}$. 
It should be pointed out that $W$ had been constructed earlier, in ref. \cite{Bandos:2002nn}, making use of an alternative approach.

Now we turn to determining the Killing supervectors of AdS$^{4|4\N}$. 
An infinitesimal isometry of AdS$^{4|4\N}$ is generated by a Killing supervector $\bm{\x}^{A} E_{A}$ which is defined to satisfy the property 
\begin{align} \label{Ads killing def}
[\bm{\x}^{A}\cD_{A} + \frac{1}{2}\l^{cd}M_{cd} + \l^{i}{}_{j}\mathbb{J}^{j}{}_{i}\,, \cD_{B}] = 0\,, 
\end{align}
for a real antisymmetric tensor $\l^{cd}(z)$. In the $\N=2$ case, $\l^{ij} = \ve^{jk}\l^{i}{}_{k}$ is symmetric, $\l^{ij} = \l^{ji}\,,$ and \eqref{Ads killing def} was solved in \cite{KT-M08}. 
Since AdS$^{4|4\N}$ is conformally related to $\N$-extended Minkowski superspace $\mathbb{M}^{4|4\N}$, see the relations \eqref{AdSBoost}, 
the supervector $\bm{\x}^{A}E_{A}$ can be decomposed with respect to the AdS basis $\{E_{A}\}$ or the flat basis $\{D_{A}\}\,,$
as 
\begin{align} \label{killing decomp}
\x = \bm{\x}^{A}E_{A} = \x^{A}D_{A}\,. 
\end{align}
Here, $\x^{A}$ are the components of a conformal Killing supervector, which generates infinitesimal superconformal transformations in $\mathbb{M}^{4|4\N}$
\begin{align}
z^{A} \longrightarrow z^{A} + \x^{A}\,, 
\end{align}
and is defined to satisfy the constraint 
\begin{align}
[\x\,, D_{\a}^{i}] \propto D_{\b}^{j}\,,
\end{align}
see, e.g., \cite{Kuzenko:2006mv,Kuzenko:1999pi}, for more details. 
With respect to the basis $\{D_{A}\}$, the components of $\x$ are 
\bsubeq \label{scf trf parameters}
\begin{align}
\tilde{\x}_{+} &= (\x_{+}^{\ad\a}) = \tilde{a} + \hf(\D + \bar{\D})\tilde{x}_{+} + \bar{K}\tilde{x}_{+} + \tilde{x}_{+}K - \tilde{x}_{+}b\tilde{x}_{+} + 4\ri\bar{\e}\q - 4\tilde{x}_{+}\eta\q
\,, \\
(\x_{i}^{\a}) &= \e + \frac{1}{2\N}\big((\N-2)\D + 2\bar{\D})\big) \q + \q K + \L\q - \q b\tilde{x}_{+} - \ri\bar{\eta}\tilde{x}_{+} - 4\q\eta\q
\,, \\
\x^{a} &= \frac{1}{2}(\x_{+}^{a} + \x_{-}^{a}) + \ri\left( \q_i\s^{a}\bar{\x}^i - \x_i \s^{a}\bar{\q}^i \right)\,, \qquad \x^{a}_{+} = -\frac{1}{2}\x_{+}^{\ad\a}(\s^{a})_{\a\ad} = \bar{\x_{-}^{a}}\,,
\end{align}
\esubeq
where the parameters $\{a\,, b\,, K\,, \bar{K}\,, \D\,, \bar{\D}\,, \e\,, \bar{\e}\,, \eta\,, \bar{\eta}\,, \L \}$ are identified with those in \eqref{scf alg}. 

Given a superconformal transformation, the compensator W transforms as 
\bsubeq \label{comp var}
\begin{align}
\d W &= \x W +  \s[\x] W\,, 
\\
 \s[\x] &= \frac{1}{\N(\N-4)}\left( (\N-2)D_{\a}^{i}\x_{i}^{\a} -  2 \bar{D}_{i}^{\ad}\bar{\x}_{\ad}^{i} \right) \,. 
\end{align}
\esubeq
Then, the problem of determining the AdS Killing supervectors proves to be equivalent to determining those conformal Killing supervectors which do not change the compensator \eqref{ads compensator}, 
\begin{align} \label{comp var = 0}
\d W = 0\,.
\end{align}
The $\N=1$ and $\N=2$ cases were worked out in \cite{Buchbinder:1998qv} and \cite{KT-M08}, respectively.
It can be shown that eq. \eqref{comp var = 0} imposes the following constraints on the transformation parameters in \eqref{scf trf parameters}
\bsubeq \label{ads killing}
\begin{align}
b^{a} &= - \frac{s^{ij}\bar{s}_{ij}}{4\N}a^{a} \,, 
\\
\eta_{\a}^{i} &= \frac{1}{2}s^{ij}\e_{\a j}\,, 
\\
s^{k(i}\L_{k}{}^{j)} &= 0\,, \label{s lambda}
\\
\D &= 0\,.
\end{align}
\esubeq
In particular, eq. \eqref{s lambda} implies 
\begin{align}
\hat{s} \L + \L^{\T} \hat{s} = 0\,, \qquad \hat{s} := (s^{ij})\,. 
\end{align}
Then, making use of \eqref{killing decomp}, one can read off the components of the AdS Killing supervector $\bm{\x}^{A}\,.$

Finally, comparing the compensator \eqref{ads compensator} with the chiral parameter $\l$ given by eq. \eqref{lambda def}, we find 
\begin{align}
\l = \re^{-\frac{1}{2}\s}\,.
\end{align}
Further, in the north chart developed in the main body, $s^{ij}$ is given by 
\begin{align} \label{coset s}
s^{ij} = 2\ri\d^{ij}\,. 
\end{align}
Inserting \eqref{coset s} into \eqref{ads killing}, we find complete agreement with the constraints derived from the supertwistor approach, eqs. \eqref{ads translation con} and \eqref{ads so(n) con}



\begin{footnotesize}

\end{footnotesize}

\end{document}